\documentstyle[12pt,epsf]{article}

\begin{document}

\font\stepone=cmr10 scaled\magstep1
\font\steptwo=cmr10 scaled\magstep2
\def\s{\smallskip}
\def\ss{\smallskip\smallskip}
\def\b{\bigskip}
\def\bb{\bigskip\bigskip}
\def\bbb{\bigskip\bigskip\bigskip}
\def\i{\item}
\def\ii{\item\item}
\def\iii{\item\item\item}
\def\sqr#1#2{{\vcenter{\vbox{\hrule height.#2pt
 \hbox{\vrule width.#2pt height#1pt \kern#1pt
 \vrule width.#2pt} \hrule height.#2pt}}}}
\def\square{\mathchoice \sqr65 \sqr65 \sqr{2.1}3 \sqr{1.5}3}
\def\utilde{{\tilde u}}
\def\operp{\hbox{${\kern+.25em{\bigcirc}
\kern-.85em\bot\kern+.85em\kern-.25em}$}}
\def\lsim{\;\raise0.3ex\hbox{$<$\kern-0.75em\raise-1.1ex\hbox{$\sim$}}\;}
\def\gsim{\;\raise0.3ex\hbox{$>$\kern-0.75em\raise-1.1ex\hbox{$\sim$}}\;}
\def\no{\noindent}
\def\r{\rightline}
\def\ce{\centerline}
\def\ve{\vfill\eject}
\def\rdots{\mathinner{\mkern1mu\raise1pt\vbox{\kern7pt\hbox{.}}\mkern2mu
 \raise4pt\hbox{.}\mkern2mu\raise7pt\hbox{.}\mkern1mu}}

\def\bu{$\bullet$}
\def\BB{$\bar B B$ }
\def\pp{$\bar p p$ }
\def\BOBO{$\bar B \bar O$BO }
\def\W{W$^{\pm}$ }
\def\ee{$e^+ e^-$ }  
\def\Z{$Z^{\circ}$ }

\rightline {UCLA/95/TEP/26}
\rightline {September 1995}
\bbb
\ce {\bf DIMENSIONAL REDUCTION AND CATALYSIS OF}
\ce {\bf DYNAMICAL SYMMETRY BREAKING BY A MAGNETIC FIELD}
\b
\ce {V. P. Gusynin,$^{1,2}$ V. A. Miransky$^{1,3}$ and I. A. Shovkovy$^1$}
\s
\ce {\it $^1$Institute for Theoretical Physics}
\ce {\it 252143 Kiev, Ukraine}
\s
\ce {\it $^2$Institute for Theoretical Physics}
\ce {\it University of Groningen, 9747 AG Groningen, The Netherlands}
\s
\ce {\it $^3$Department of Physics}
\ce {\it University of California, Los Angeles, CA  90095-1547, USA}
\bbb
\no {\bf ABSTRACT.}  It is shown that a constant magnetic field in 3+1
and 2+1 dimensions is a strong catalyst of dynamical chiral
symmetry breaking, leading to the generation of a fermion dynamical mass even
at the weakest attractive interaction between fermions.  The essence
of this effect is the dimensional reduction $D\to D-2$ in the
dynamics of fermion pairing in a magnetic field.  The effect is
illustrated in the Nambu--Jona--Lasinio (NJL) model and QED.  In the
NJL model in a magnetic field, the low--energy effective action 
and the spectrum of long wavelength collective excitations are
derived.  In QED (in ladder and improved ladder 
approximations) the dynamical mass of fermions (energy gap in the 
fermion spectrum) is determined. Possible 
applications of this effect and its extension to inhomogeneous 
field configurations are discussed.
\ve
\ce {\bf 1.  INTRODUCTION}
\b

At present there are only a few firmly established non-perturbative
phenomena in 2+1 and, especially, $(3+1)$--dimensional field theories. 
In this paper, we will establish and describe one more such phenomenon:
dynamical chiral symmetry breaking by a magnetic field.

The problem of fermions in a constant magnetic field had been
considered by Schwinger long ago \cite{1}.  In that classical work,
while the interaction with the external magnetic field was considered
in all orders in the coupling constant, quantum dynamics was
treated perturbatively.  There is no spontaneous chiral symmetry
breaking in this approximation.  In this paper we will reconsider
this problem, treating quantum dynamics non-perturbatively.  We will
show that in 3+1 and 2+1 dimensions, a constant magnetic field is a
strong catalyst of dynamical chiral symmetry breaking, leading to
generating a fermion mass even at the weakest attractive interaction 
between fermions.  We stress that this effect is universal, i.e.
model independent.

The essence of this effect is the dimensional reduction $D\to D-2$
in the dynamics of fermion pairing in a magnetic field: while at
$D=2+1$ the reduction is $2+1\to 0+1$, at $D=3+1$ it is
$3+1\to 1+1$.  The physical reason of this reduction is the
fact that the motion of charged particles is restricted in directions
perpendicular to the magnetic field.

Since the case of 2+1 dimensions has been already considered in detail
in Ref. \cite{2}, the emphasis in this paper will be on $(3+1)$--dimensional
field theories.  However, it will be instructive to compare the
dynamics in 2+1 and 3+1 dimensions.

As concrete models for the quantum dynamics, we consider the
Nambu-Jona-Lasinio (NJL) model \cite{3} and QED.  We will show that
the dynamics of the lowest Landau level (LLL) plays the crucial
role in catalyzing spontaneous chiral symmetry breaking.  Actually,
we will see that the LLL plays here the role similar to that of the
Fermi surface in the BCS theory of superconductivity \cite{4}.

As we shall show in this paper, the dimensional reduction $D\to D-2$
is reflected in the structure of the equation describing the 
Nambu--Goldstone (NG) modes in a magnetic field.  In Euclidean space,
for weakly interacting fermions, it has the form of a two-dimensional
(one--dimensional) Schr\"odinger equation at $D=3+1~(D=2+1)$:
\begin{eqnarray}
\bigl(-\Delta+m^2_{\rm dyn} + V({\bf r})\bigr) \Psi({\bf r})=0~.
\label{1}
\end{eqnarray}
\no Here $\Psi({\bf r})$ is expressed through the Bethe--Salpeter (BS)
function of NG bosons,
\begin{eqnarray*}
\Delta={\partial^2\over \partial x_3^2}+{\partial^2\over \partial x^2_4}
\end{eqnarray*}
\no (the magnetic field is in the $+x_3$ direction, $x_4=it$) for
$D=3+1$, and $\Delta={\partial^2\over \partial x_3^2}$, $x_3=it$, for
$D=2+1$.  The attractive potential $V(\bf r)$ is model dependent.
In the NJL model (both at $D=2+1$ and $D=3+1$), $V({\bf r})$ is a
$\delta$--like potential.  In $(3+1)$--dimensional ladder QED, the potential
$V({\bf r})$ is
\begin{eqnarray}
V({\bf r})={\alpha\over \pi\ell^2} \exp
\biggl({r^2\over 2\ell^2}\biggr) Ei\biggl(-{r^2\over 2\ell^2}\biggr),
\quad
r^2=x_3^2+x_4^2~, \label{2}
\end{eqnarray}
\no where $Ei(x)=-\int^\infty_{-x} dt \exp(-t)/t$ is the integral
exponential function \cite{5}, $\alpha={e^2\over 4\pi}$ is the renormalized
coupling constant and $\ell\equiv |eB|^{-1/2}$ is the magnetic length.

Since $-m^2_{\rm dyn}$ plays the role of energy $E$ in this
equation and $V({\bf r})$ is an attractive potential, the problem is
reduced to finding the spectrum of bound states (with 
$E=-m^2_{\rm dyn}<0$) of the Schr\"odinger equation with such a
potential.  More precisely, since only the largest possible value 
of $m^2_{\rm dyn}$ defines the stable vacuum \cite{6}, we need to find
the lowest eigenvalue of $E$.  For this purpose, we can use results
proved in the literature for the one--dimensional $(d=1)$ and two--dimensional
$(d=2)$ Schr\"odinger equation \cite{7}.  These results ensure that there is
at least one bound state for an attractive potential for $d=1$ and $d=2$.
The energy of the lowest level $E$ has the form:
\begin{eqnarray}
E(\lambda)=-m^2_{\rm dyn}(\lambda)=-|eB|f(\lambda)~, \label{3}
\end{eqnarray}
\no where $\lambda$ is a coupling constant ($\lambda$ is $\lambda=G$ 
in the NJL model and $\lambda=\alpha$ in QED).  While for $d=1$,
$f(\lambda)$ is an analytic function of $\lambda$ at $\lambda=0$, for
$d=2$, it is non--analytic at $\lambda=0$.  Actually we find that,
as $G\to 0$,
\begin{eqnarray}
m^2_{\rm dyn}=|eB|{N_c^2G^2|eB|\over 4\pi^2}~, \label{4}
\end{eqnarray}
\no where $N_c$ is the number of fermion colors, in $(2+1)$--dimensional
NJL model \cite{2}, and
\begin{eqnarray}
m^2_{\rm dyn}={|eB|\over \pi}
\exp\biggl(-{4\pi^2(1-g)\over |eB|N_c G}\biggr)~, \label{5}
\end{eqnarray}
\no where $g\equiv N_cG\Lambda^2/(4\pi^2)$, in $(3+1)$-dimensional
NJL model.  In $(3+1)$-dimensional ladder QED, $m_{\rm dyn}$ is 
\begin{eqnarray}
m_{\rm dyn} = C\sqrt{|eB|}\exp \biggl[-{\pi\over2}\biggl({\pi\over2\alpha}
\biggr)^{1/2}\biggr]~, \label{6}
\end{eqnarray}
\no where the constant $C$ is of order one and $\alpha$ is the
renormalized coupling constant. As we will show below, this 
expression for $m_{\rm dyn}$ in QED is gauge invariant.

It is important that, as we shall show below, an infrared dynamics,
with a weak QED coupling, is
responsible for spontaneous symmetry breaking in QED in a magnetic
field. This suggests that the ladder
approximation can be reliable in this problem. However, because of
the $(1+1)$--dimensional form of the fermion propagator in the
infrared region, there may be also relevant higher order
contributions. As we shall show in this paper, there are indeed
relevant one--loop  contributions in the
photon propagator. Taking into account these contributions (that
corresponds to the so called improved ladder approximation), we get
the expression for $m_{\rm dyn}$ of the form (\ref{6}) with
$\alpha\to\alpha/2$. We  shall discuss the physics underlying this
effect and the general status of the
validity of the expansion in $\alpha$ in the infrared
region in QED in a magnetic field in Sec.~9. 

As we will discuss in this paper, there may exist interesting 
applications of the phenomenon of chiral symmetry breaking by a
magnetic field: in planar condensed matter systems, 
in cosmology, in the
interpretation of the heavy-ion scattering experiments, and for
understanding of the structure of the QCD vacuum.
We will also discuss an extension of these results to inhomogeneous
field configurations.

The paper is organized as follows.  In Section 2 we consider the problem
of a free relativistic fermion in a magnetic field in 3+1 and 2+1
dimensions.  We show that the roots of the fact that a magnetic field
is a strong catalyst of dynamical chiral symmetry breaking are actually
in this problem.  In Section 3 we show that the dimensional reduction
$3+1 \to 1+1$ ($2+1\to 0+1$) in the dynamics of fermion
pairing in a magnetic field is consistent with spontaneous chiral 
symmetry breaking and does not contradict the Mermin-Wagner-Coleman
theorem forbidding the spontaneous breakdown of continuous symmetries
in less than 2+1 dimensions.
In Sections 4-8 we study the NJL model in a magnetic
field in 3+1 dimensions.  We derive the low-energy effective action
and determine the spectrum of long wavelength collective excitations
in this model.  We also compare these results with those in the
$(2+1)$--dimensional NJL model \cite{2}.  
In Section 9 we study dynamical chiral 
symmetry breaking in QED in a magnetic field.  In Section 10 we 
summarize the main results of the paper and discuss possible applications
of this effect.  In Appendix A some useful formulas and relations are 
derived.  In Appendix B the reliability of the $1/N_c$ expansion in the 
NJL model in a magnetic field is discussed. In Appendix C we analyze
the Bethe-Salpeter equation for Nambu-Goldstone bosons in QED in a
magnetic field.
\ve
\ce {\bf 2. FERMIONS IN A CONSTANT MAGNETIC FIELD}
\b

In this section we will discuss the problem of relativistic fermions
in a magnetic field in 3+1 dimensions and compare it with the same
problem in 2+1 dimensions.  We will show that the roots of the fact that
a magnetic field is a strong catalyst of chiral symmetry breaking are
actually in this dynamics.

The Lagrangian density in the problem of a relativistic fermion in a
constant magnetic field $B$ takes the form
\begin{eqnarray}
{\cal{L}}={1\over 2}\bigl[\bar\psi,~(i\gamma^\mu D_\mu-m)\psi\bigr]~,
\quad
\mu=0,1,2,3, \label{7}
\end{eqnarray}
\no where the covariant derivative is
\begin{eqnarray}
D_\mu=\partial_\mu-ie A_\mu^{\rm ext}~. \label{8}
\end{eqnarray}
\no We will use two gauges in this paper:
\begin{eqnarray}
A_\mu^{\rm ext} =-\delta_{\mu 1}Bx_2 \label{9}
\end{eqnarray}
\no (the Landau gauge) and 
\begin{eqnarray}
A_\mu^{\rm ext} = -{1\over 2} \delta_{\mu 1}Bx_2+{1\over 2}
\delta_{\mu 2}Bx_1 \label{10}
\end{eqnarray}
\no (the symmetric gauge).  The magnetic field is in the $+x_3$
direction.

The energy spectrum of fermions is \cite{8}:
\begin{eqnarray}
E_n(k_3)=\pm \sqrt{m^2+2|eB|n+k_3^2}~, \quad
n=0,1,2,~\ldots \label{11}
\end{eqnarray}
\no (the Landau levels).  Each Landau level is degenerate: at each
value of the momentum $k_3$, the number of states is
\begin{eqnarray*}
dN_0=S_{12}L_3 {|eB|\over 2\pi}
{dk_3\over 2\pi}
\end{eqnarray*}
\no at n=0, and 
\begin{eqnarray*}
dN_n=S_{12}L_3 {|eB|\over \pi}{dk_3\over 2\pi}
\end{eqnarray*}
\no at $n\geq 1$ (where $L_3$ is the size in the $x_3$-direction and
$S_{12}$ is the square in the $x_1x_2$-plane).  In the Landau gauge
(\ref{9}),
the degeneracy is connected with the momentum $k_1$ which at the same
time coincides with the $x_2$ coordinate of the center of a fermion orbit
in the $x_1x_2$-plane \cite{8}.  In the symmetric gauge (\ref{10}), the degeneracy is
connected with the angular momentum $J_{12}$ in the $x_1x_2$-plane (for a
recent review see Ref. \cite{9}).

As the fermion mass $m$ goes to zero, there is no energy gap between the
vacuum and the lowest Landau level (LLL) with $n=0$.  The density of
the number of states of fermions on the energy surface with $E_0=0$ is
\begin{eqnarray}
\nu_0=V^{-1} {dN_0\over dE_0}\biggl|_{E_0=0}=
S_{12}^{-1} L_3^{-1}{dN_0\over dE_0}\biggl|_{E_0=0}=
{|eB|\over 4\pi^2}~, \label{12}
\end{eqnarray}
\no where $E_0=|k_3|$ and $dN_0=V{|eB|\over 2\pi}{dk_3\over 2\pi}$
(here $V=S_{12}L_3$ is the volume of the system).  We will see that
the dynamics of the LLL plays the crucial role in catalyzing spontaneous
chiral symmetry breaking.  In particular, the density $\nu_0$ plays the
same role here as the density of states on the Fermi surface $\nu_F$
in the theory of superconductivity \cite{4}.

The important point is that the dynamics of the LLL is essentially
(1+1)-dim\-en\-sion\-al.  In order to see this, let us consider the fermion
propagator in a magnetic field.  It was calculated by Schwinger \cite{1}
and has the following form both in the gauge (\ref{9}) and the gauge
(\ref{10}):
\begin{eqnarray}
S(x,y)=\exp\biggl[{ie\over 2}(x-y)^\mu A_\mu^{\rm ext}(x+y)\biggr]
\tilde S(x-y)~, \label{13}
\end{eqnarray}
\no where the Fourier transform of $\tilde S$ is
\begin{eqnarray}
\tilde S(k) &=& \int^\infty_0 
ds\exp\biggl[is\bigl(k_0^2-k_3^2-{\bf k^2}_{\perp}
{\tan (eBs)\over eBs} - m^2\bigr)\biggr] \nonumber\\ 
&&\hspace{-10mm}\cdot[\bigl(k^0\gamma^0-k^3
\gamma^3+m)(1+\gamma^1\gamma^2 \tan(eBs)\bigr)
-{\bf k}_{\perp}{\mbox{\boldmath$\gamma$}}_{\perp}(1+\tan^2(eBs)\bigr)\bigr]  
\label{14}
\end{eqnarray}
\no (here $\bf k_{\perp}=(k_1,k_2),~
{\mbox{\boldmath $\gamma$}}_{\perp}=(\gamma_1,\gamma_2)$).
Then by using the identity $i\tan(x)=   \break
1-2\exp(-2ix)/[1+\exp(-2ix)]$ and
the relation \cite{5}
\begin{eqnarray}
(1-z)^{-(\alpha+1)}\exp\biggl({xz\over z-1}\biggr)=\sum^\infty_{n=0}
L^\alpha_n(x)z^n~, \label{15}
\end{eqnarray}
\no where $L_n^\alpha(x)$ are the generalized Laguerre polynomials,
the propagator $\tilde S(k)$ can be decomposed over the Landau poles \cite{10}:
\begin{eqnarray}
\tilde S(k)=i \exp\biggl(-{{\bf k}_{\perp}^2\over |eB|}\biggr) 
\sum^\infty_{n=0}
(-1)^n {D_n(eB,k)\over k_0^2-k_3^2-m^2-2|eB|n} \label{16}
\end{eqnarray}
\no with
\begin{eqnarray}
D_n(eB,k)&=&(k^0\gamma^0-k^3\gamma^3+m)
\biggl[\bigl(1-i\gamma^1\gamma^2 {\rm sign}(eB)\bigr)L_n
\biggl(2{{\bf k}_{\perp}^2\over |eB|}\biggr) \nonumber\\
&&\hspace{-25mm}-\bigl(1+i\gamma^1\gamma^2{\rm sign}(eB)\bigr)L_{n-1}
\biggl(2{{\bf k}_{\perp}^2\over |eB|}\biggr)\biggr]
+4(k^1 \gamma^1+k^2\gamma^2)L^1_{n-1}
\biggl(2{{\bf k}_{\perp}^2\over |eB|}\biggr)\, , 
\end{eqnarray}  

\no where $L_n \equiv L_n^0$ and $L_{-1}^\alpha=0$ by definition.
The LLL pole is
\begin{eqnarray}
\tilde S^{(0)}(k)=i~\exp\biggl(-{{\bf k}_{\perp}^2\over |eB|}\biggr)
{k^0\gamma^0-k^3\gamma^3+m\over k_0^2-k_3^2-m^2}
\bigl(1-i\gamma^1\gamma^2 {\rm sign}(eB)\bigr)~. \label{18}
\end{eqnarray}
\no This equation clearly demonstrates the $(1+1)$--dimensional
character of the LLL dynamics in the infrared region, with
${\bf k}_{\perp}^2 \ll |eB|$.  Since at $m^2,k_0^2,k_3^2,
{\bf k}_{\perp}^2 \ll |eB|$ the LLL pole dominates in the fermion
propagator, one concludes that the dimensional reduction 
$3+1\to 1+1$ takes place for the infrared dynamics
in a strong (with $|eB| \gg m^2$) magnetic field.  It is clear
that such a dimensional reduction reflects the fact that the
motion of charged particles is restricted in directions perpendicular
to the magnetic field.

The LLL dominance can, in particular, be seen in the calculation of
the chiral condensate:
\begin{eqnarray}
\langle 0|\bar\psi\psi|0\rangle &=&
-\lim_{x\to y} {\rm tr}~S(x,y)=-{i\over (2\pi)^4} {\rm tr}~\int
d^4k~\tilde S_E(k) \nonumber\\
&&\hspace{-30mm}= -{4m\over (2\pi)^4}\int~d^4k \int^\infty_{1/\Lambda^2} ds
\exp\biggl[-s\biggl(m^2+k^2_4+k^2_3+{\bf k}^2_{\perp}
{\tanh(eBs)\over eBs}\biggr)\biggr] \nonumber\\
&&\hspace{-30mm}= -|eB| {m\over 4\pi^2} \int^\infty_{1/\Lambda^2} 
{ds\over s}
e^{-sm^2} \coth(|eBs|)
\longrightarrow_{_{_{_{\hspace{-7mm}
{{ m\rightarrow}{ 0}}}}}}
-|eB|{m\over 4\pi^2}
\biggl(\ln{\Lambda^2\over m^2}+O(m^0)\biggr) , \label{19}
\end{eqnarray}

\no where $\tilde S_E(k)$ is the image of $\tilde S(k)$ in
Euclidean space and $\Lambda$ is an ultraviolet cutoff.  
As it is clear from Eqs. (\ref{16}) and (\ref{18}), the logarithmic
singularity in the condensate appears due to the LLL dynamics.

The above consideration suggests that there is a universal mechanism
for enhancing the generation of fermion masses by a strong magnetic
field in 3+1 dimensions: the fermion pairing takes place essentially
for fermions at the LLL and this pairing dynamics is
$(1+1)$--dimensional
(and therefore strong) in the infrared region.  This in turn suggests
that in a magnetic field, spontaneous chiral symmetry breaking takes
place even at the weakest attractive interaction between fermions in
3+1 dimensions.  In this paper we will show the existence of this
effect in the NJL model and QED.

In conclusion, let us compare the dynamics in a magnetic field in 3+1
dimensions with that in 2+1 dimensions \cite{2}.  In 2+1 dimensions, we
consider the four--component fermions \cite{11}, connected with a 
four--dimensional (reducible) representation of Dirac's matrices
\begin{eqnarray}
\gamma^0=\left(\matrix{\sigma_3 & 0\cr 0 & -\sigma_3\cr}\right)~, \quad
\gamma^1=\left(\matrix{i\sigma_1 & 0\cr 0 & -i\sigma_1\cr}\right)~, \quad
\gamma^2=\left(\matrix{i\sigma_2 & 0\cr 0 & -i\sigma_2\cr}\right)~.
\label{20}
\end{eqnarray}
\no The Lagrangian density is:
\begin{eqnarray}
{\cal{L}}={1\over 2}\bigl[\bar\psi,~(i\gamma^\mu D_\mu-m)
\psi\bigr]~, \label{21}
\end{eqnarray}
\no where $D_\mu=\partial_\mu-ieA_\mu^{\rm ext}$ with 
$A_\mu^{\rm ext}=(0,-Bx_2,0)$ or $A_\mu^{\rm ext}=
\bigl(0,-{B\over 2} x_2,{B\over 2} x_1\bigr)$.  At $m=0$, this
Lagrangian density is invariant under the flavor (chiral) $U(2)$
transformations with the generators
\begin{eqnarray}
T_0=I,~ T_1=\gamma_5,~ T_2={1\over i}\gamma^3,~T_3=\gamma^3\gamma^5
\label{22}
\end{eqnarray}
\no where $\gamma^5=i\gamma^0\gamma^1\gamma^2\gamma^3=i
\left(\matrix{0 & I\cr -I & 0\cr}\right)$.  The mass term breaks
this symmetry down to the $U(1)\times U(1)$ with the generators
$T_0$ and $T_3$.  The fermion propagator in 2+1-dimensions is \cite{2}:
\begin{eqnarray}
S(x,y)=\exp\biggl({ie\over 2}(x-y)^\mu A_\mu^{\rm ext}(x+y)\biggr)
\tilde S(x-y)~, \label{23}
\end{eqnarray}
\no where the Fourier transform $\tilde S(k)$ of $\tilde S(x)$ is
\begin{eqnarray}
\tilde S(k) &=& \int^\infty_{1/\Lambda^2} 
ds \exp\biggl[-ism^2+isk_0^2
-is{\bf k}^2 {\tan(eBs)\over eBs}\biggr] \nonumber\\
&& \cdot \bigl[(k^\mu\gamma_\mu+m+(k^2\gamma^1-k^1\gamma^2)
\tan(eBs))(1+\gamma^1\gamma^2\tan (eBs))\bigr]~. 
\end{eqnarray} 
\no The decomposition $\tilde S(k)$ over the Landau level poles now
takes the following form:
\begin{eqnarray}
\tilde S(k)=i~\exp\biggl(-{{\bf k}^2\over |eB|}\biggr)
\sum^\infty_{n=0} (-1)^n
{D_n(eB,k)\over k_0^2-m^2-2|eB|n} \label{25}
\end{eqnarray}
\no with
\begin{eqnarray}
D_n(eB,k)&=&(k^0\gamma^0+m)\biggl[
\bigl(1-i\gamma^1\gamma^2{\rm sign}
(eB)\bigr)L_n\biggl(2{{\bf k}^2\over |eB|}\biggr) \nonumber\\
&&\hspace{-30mm}-\bigl(1+i\gamma^1\gamma^2{\rm sign}(eB)\bigr)L_{n-1}
\biggl(2{{\bf k}^2\over |eB|}\biggr)\biggr]
+4(k^1\gamma^1+k^2\gamma^2)L^1_{n-1}
\biggl(2{{\bf k}^2\over |eB|}\biggr)~. 
\end{eqnarray}
\no Then Eq. (\ref{25}) implies that as $m\to 0$, the condensate
$\langle 0|\bar\psi\psi|0\rangle$ remains non-zero due to the LLL:
\begin{eqnarray}
\langle 0|\bar\psi\psi|0\rangle =-\lim_{m\to 0}
{m\over (2\pi)^3} \int d^3k
{\exp(-{\bf k}^2/|eB|)\over k_3^2+m^2} = -{|eB|\over 2\pi}
\label{27}
\end{eqnarray}
\no (for concreteness, we consider $m\geq 0$).  The appearance of
the condensate in the flavor (chiral) limit, $m\to 0$, signals
the spontaneous breakdown of $U(2)$ to $U(1)\times U(1)$ \cite{2}.
As we will discuss in Section 5, this in turn provides the analyticity 
of the dynamical mass $m_{\rm dyn}$ as a function of the coupling constant
$G$ at $G=0$ in the $(2+1)$--dimensional NJL model.
\ve

\ce {\bf 3. IS THE DIMENSIONAL REDUCTION 3+1 $\to$ 1+1 (2+1 
$\to$ 0+1)} 
\ce {\bf CONSISTENT WITH SPONTANEOUS CHIRAL} 
\ce {\bf SYMMETRY BREAKING?}
\b

In this section we consider the question whether the dimensional
reduction 3+1 $\to$ 1+1 (2+1 $\to$ 0+1) in the dynamics
of the fermion pairing in a magnetic field is consistent with spontaneous
chiral symmetry breaking.  This question occurs naturaly since, due to the
Mermin-Wagner-Coleman (MWC) theorem \cite{12}, there cannot be spontaneous 
breakdown of continuous symmetries at $D=1+1$ and $D=0+1$.  The MWC theorem
is based on the fact that gapless Nambu-Goldstone (NG) bosons cannot exist
in dimensions less than 2+1.  This is in particular reflected in that the
$(1+1)$--dimensional propagator of would be NG bosons would lead to infrared
divergences in perturbation theory (as indeed happens in the $1/N_c$ 
expansion in the $(1+1)$--dimensional Gross--Neveu model with a continuous 
symmetry \cite{13}).

However, the MWC theorem is not applicable to the present problem.  The
central point is that the condensate $\langle 0|\bar\psi\psi|0\rangle$
and the NG modes are {\bf neutral} in this problem and the dimensional
reduction in a magnetic field does not affect the dynamics of the center
of mass of {\bf neutral} excitations.  Indeed, the dimensional reduction
$D\to D-2$ in the fermion propagator, in the infrared region, 
reflects the fact that the motion of {\bf charged} particles is 
restricted in the directions perpendicular to the magnetic field.  Since
there is no such restriction for the motion of the center of mass of 
neutral excitations, their propagators have $D$--dimensional form in the
infrared region (since the structure of excitations is irrelevant at long
distances, this is correct both for elementary and composite neutral
exci-\break tations).
\footnote{The Lorentz invariance is broken by a
magnetic field in this problem.  By the $D$--dimensional form, we
understand that the denominator of the propagator depends on energy
and all the components of momentum.  That is, for $D=3+1,~{\cal{D}}(P)\sim
(P_0^2-C_{\perp} {\bf P}_{\perp}^2-C_3P_3^2)^{-1}$ with $C_{\perp},C_3
\not= 0$.}
This fact will be shown for neutral bound states in the NJL model
in a magnetic field, in the $1/N_c$ expansion, in Section 6 and Appendix B.
Since, besides that, the propagator of {\bf massive} fermions is, though
$(D-2)$--dimensional, nonsingular at small momenta, the infrared dynamics is
soft in a magnetic field, and spontaneous chiral symmetry breaking is not
washed out by the interactions, as happens, for example, in the 
$(1+1)$--dimensional Gross--Neveu model \cite{13}.

This point is intimately connected with the status of the space-translation
symmetry in a constant magnetic field.  In the gauge (\ref{9}), the 
translation symmetry along the $x_2$ direction is broken; in the gauge
(\ref{10}), the translation symmetry along both the $x_1$ and $x_2$ 
directions is broken.  However, for neutral states, all the components
of the momentum of their center of mass are conserved quantum numbers
(this property is gauge invariant) \cite{14}.  In order to show this, let us
introduce the following operators (generators of the group of magnetic
translations) describing the translations in first quantized theory:
\begin{eqnarray}
\hat P_{x_1}={1\over i}{\partial\over \partial x_1}~, \quad
\hat P_{x_2}={1\over i}{\partial\over \partial x_2}+\hat Q Bx_1~, \quad
\hat P_{x_3}={1\over i}{\partial\over \partial x_3} \label{28}
\end{eqnarray}
\no in gauge (\ref{9}), and
\begin{eqnarray}
\hat P_{x_1}={1\over i}{\partial \over \partial x_1}-
{\hat Q\over 2} Bx_2~, \quad
\hat P_{x_2} = {1\over i}{\partial\over \partial x_2}+
{\hat Q\over 2} Bx_1~, \quad
\hat P_{x_3}={1\over i}{\partial\over \partial x_3}  \label{29}
\end{eqnarray}
\no in gauge (\ref{10}) ($\hat Q$ is the charge operator).  One can easily
check that these operators commute with the Hamiltonian of the Dirac
equation in a constant magnetic field.  Also, we get:
\begin{eqnarray}
\bigl[\hat P_{x_1},\hat P_{x_2}\bigr] = {1\over i}
\hat QB~, \quad
\bigl[\hat P_{x_1},\hat P_{x_3}\bigr] = 
\bigl[\hat P_{x_2},\hat P_{x_3}\bigr] = 0~. \label{30}
\end{eqnarray}
\no Therefore all the commutators equal zero for neutral states, and
the momentum ${\bf P}=(P_1,P_2,P_3)$ can be used to describe the
dynamics of the center of mass of neutral states.
\ve

\ce {\bf 4. THE NAMBU-JONA-LASINIO MODEL IN A MAGNETIC FIELD:}
\ce {\bf GENERAL CONSIDERATION}
\b

In this and the next four sections we shall consider the NJL model in a
magnetic field.  This model gives a clear illustration of the general
fact that a constant magnetic field is a strong catalyst for the generation
of a fermion dynamical mass.

Let us consider the $(3+1)$--dimensional NJL model with the $U_{\rm L}(1)\times
U_{\rm R}(1)$ chiral symmetry:
\begin{eqnarray}
{\cal{L}}={1\over 2}\bigl[\bar\psi,(i\gamma^\mu D_\mu) \psi\bigr]+
{G\over 2}\bigl[(\bar\psi\psi)^2+
(\bar\psi i\gamma^5\psi)^2\bigr]~, \label{31}
\end{eqnarray}
\no where $D_\mu$ is the covariant derivative (\ref{8}) and fermion fields
carry an additional ``color" index $\alpha=1,2,\ldots,N_c$.  The theory
is equivalent to the theory with the Lagrangian density
\begin{eqnarray}
{\cal{L}}={1\over 2}\bigl[\bar\psi,(i\gamma^\mu D_\mu)\psi\bigr]-
\bar\psi(\sigma+i\gamma^5\pi)\psi-{1\over 2G}
(\sigma^2+\pi^2)~. \label{32}
\end{eqnarray}
\no The Euler-Lagrange equations for the auxiliary fields $\sigma$
and $\pi$ take the form of constraints:
\begin{eqnarray}
\sigma=-G(\bar\psi\psi)~, \quad
\pi=-G(\bar\psi i\gamma^5\psi)~, \label{33}
\end{eqnarray}
\no and the Lagrangian density (\ref{32}) reproduces Eq. (\ref{31}) upon application
of the constraints (\ref{33}).

The effective action for the composite fields is expressed through the
path integral over fermions:
\begin{eqnarray}
\Gamma(\sigma,\pi)=\tilde\Gamma(\sigma,\pi)-{1\over 2G} \int
d^4x (\sigma^2+\pi^2)~, \label{34}
\end{eqnarray}
\begin{eqnarray}
\exp(i\tilde\Gamma)&=&\int [d\psi][d\bar\psi]
\exp
\left\{ 
{i\over 2}\int d^4x
\bigl[
\bar\psi,\{i\gamma^\mu D_\mu-(\sigma+i\gamma^5\pi)\}\psi
\bigr]
\right\}  \nonumber\\
&=&
\exp
\left\{
{\rm Tr}\,{\rm Ln}
\biggl[
i\gamma^\mu D_\mu-\bigl( \sigma+i\gamma^5\pi \bigr)
\biggr]
\right\},  
\end{eqnarray}
i.e.
\begin{eqnarray}
\tilde\Gamma(\sigma,\pi)=-i{\rm Tr}\,{\rm Ln}\bigl[i\gamma^\mu D_\mu-
(\sigma+i\gamma^5\pi)\bigr]~. \label{36}
\end{eqnarray}
\no As $N_c\to \infty$, the path integral over the composite
fields $\sigma$ and $\pi$ is dominated by the stationary points of
the action: $\delta\Gamma/\delta\sigma=\delta\Gamma/\delta\pi=0$.
We will analyze the dynamics in this limit by using the expansion of
the action $\Gamma$ in powers of derivatives of the composite fields.

Is the $1/N_c$ expansion reliable in this problem?  The answer to this
question is ``yes".  It is connected with the fact, already discussed
in the previous section, that the dimensional reduction 3+1 $\to$
1+1 by a magnetic field does not affect the dynamics of the center of
mass of the NG bosons.  If the reduction affected it, the $1/N_c$
perturbative expansion would be unreliable.  In particular, the
contribution of the NG modes in the gap equation, in next--to--leading order
in $1/N_c$, would lead to infrared divergences (as happens in the
$(1+1)$--dimensional Gross--Neveu model with a continuous chiral symmetry \cite{13}).
This is not the case here.  Actually, as we will show in Appendix B,
the next-to-leading order in $1/N_c$ yields small corrections to the
whole dynamics at sufficiently large values of $N_c$.
\ve

\ce {\bf 5. THE NJL MODEL IN A MAGNETIC FIELD:} 
\ce {\bf THE EFFECTIVE POTENTIAL}
\b

We begin the calculation of the effective action $\Gamma$ by calculating
the effective potential $V$.  Since $V$ depends only on the
$U_{\rm L}(1)\times U_{\rm R}(1)$-invariant $\rho^2=\sigma^2+\pi^2$,
it is sufficient to consider a configuration with $\pi=0$ and $\sigma$ 
independent of $x$.  So now $\tilde\Gamma(\sigma)$ in Eq. (\ref{36}) is:
\begin{eqnarray}
\tilde\Gamma=-i~
{\rm Tr}\,{\rm Ln} (i\hat D-\sigma)=
-i~\ln\,{\rm Det}(i\hat D-\sigma)~, \label{37}
\end{eqnarray}
\no where $\hat D\equiv \gamma^\mu D_\mu$.  Since
\begin{eqnarray}
{\rm Det}(i\hat D-\sigma)={\rm Det}(\gamma^5(i\hat D-\sigma)\gamma^5)=
{\rm Det}(-i\hat D-\sigma)~, \label{38}
\end{eqnarray}
\no we find that
\begin{eqnarray}
\tilde\Gamma(\sigma)=-{i\over 2}{\rm Tr}
\bigl[\,{\rm Ln}(i\hat D-\sigma)+
{\rm Ln}(-i\hat D-\sigma)\bigr] 
=-{i\over 2}{\rm Tr}\,{\rm Ln}(\hat D^2+\sigma^2)\, . 
\end{eqnarray}
\no Therefore $\tilde\Gamma(\sigma)$ can be expressed through the 
following integral over the proper time $s$:
\begin{eqnarray}
\tilde\Gamma(\sigma)=-{i\over 2}{\rm Tr}\,{\rm Ln}(\hat D^2+\sigma^2)
={i\over 2} \int d^4x \int^\infty_0 {ds\over s} {\rm tr}
\langle x|e^{-is(\hat D^2+\sigma^2)}|x\rangle \label{40}
\end{eqnarray}
\no where
\begin{eqnarray}
\hat D^2=D_\mu D^\mu-{ie\over 2}
\gamma^\mu\gamma^\nu F_{\mu\nu}^{\rm ext}=
D_\mu D^\mu+ie\gamma^1\gamma^2 B~. \label{41}
\end{eqnarray}
\no The matrix element $\langle x|e^{is(\hat D^2+\sigma^2)}|y\rangle$
was calculated by Schwinger \cite{1}:
\begin{eqnarray}
\langle x|e^{-is(\hat D^2+\sigma^2)}|y\rangle  &=&
e^{-is\sigma^2}\langle x|e^{-isD_\mu
D^\mu}|y\rangle 
\bigl[\cos(eBs)+\gamma^1\gamma^2\cdot \sin(eBs)\bigr] \nonumber\\
&=& {-i\over (4\pi s)^2} e^{-i(s\sigma^2-S_{c\ell})}
\bigl[ eBs\cot(eBs)+\gamma^1\gamma^2 eBs\bigr]~, \label{42}
\end{eqnarray}
\no where
\begin{eqnarray}
S_{c\ell}&=&e \int^x_y A_\lambda^{\rm ext} dz^\lambda-
{1\over 4s}(x-y)_\nu 
\bigl(g^{\nu\mu}+\frac{(F^2_{\rm ext})^{\nu\mu}}{B^2} \nonumber\\
&&\cdot \bigl[1-eBs \cot(eBs)\bigr]\bigr)(x-y)_\mu~ .  \label{43}
\end{eqnarray}
\no Here the integral $\int^x_y A_\lambda^{\rm ext} dz^\lambda$ is
taken along a straight line.  Substituting expression (\ref{42}) into
Eq. (\ref{40}), we find
\begin{eqnarray}
\tilde\Gamma(\sigma)={N_c\over 8\pi^2 } \int d^4x \int^\infty_0
{ds\over s^2} e^{-is\sigma^2} eB \cot(eBs)~. \label{44}
\end{eqnarray}
\no Therefore the effective potential is
\begin{eqnarray}
V(\rho)={\rho^2\over 2G}+\tilde V(\rho)=
{\rho^2\over 2G}+{N_c\over 8\pi^2} \int^\infty_{1/\Lambda^2}
{ds\over s^2} e^{-s\rho^2} eB\coth (eBs)~, \label{45}
\end{eqnarray}
\no where now the ultraviolet cutoff has been explicitly introduced.

By using the integral representation for the generalized Riemann
zeta function $\zeta$ \cite{5},
\begin{eqnarray}
\int^\infty_0 ds s^{\mu-1} e^{-\beta s} \coth s=
\Gamma(\mu)\biggl[2^{1-\mu}\zeta\biggl(\mu,~{\beta\over 2}\biggr)
-\beta^{-\mu}\biggr]~, \label{46}
\end{eqnarray}
\no which is valid at $\mu>1$, and analytically continuing this
representation to negative values of $\mu$, we can rewrite Eq.
(\ref{45}) as
\begin{eqnarray}
V(\rho)&=&{\rho^2\over 2G}+{N_c\over 8\pi^2}
\biggl[{\Lambda^4\over 2}+{1\over 3\ell^4} \ln(\Lambda\ell)^2+
{1-\gamma-\ln 2\over 3\ell^4} \nonumber\\
&-& (\rho\Lambda)^2+{\rho^4\over 2}\ln(\Lambda\ell)^2+{\rho^4\over 2}
(1-\gamma-\ln 2)+{\rho^2\over \ell^2}\ln{\rho^2\ell^2\over 2} \nonumber\\
&-& {4\over \ell^4} \zeta^\prime\biggl(-1,{\rho^2\ell^2\over 2}
+1\biggr)\biggr] + O(1/\Lambda)~, \label{47}
\end{eqnarray}
\no where the magnetic length $\ell\equiv |eB|^{-1/2},~
\zeta^\prime(-1,x)={d\over d\nu} \zeta(\nu,x)\bigl|_{\nu=-1}$, and
$\gamma\simeq 0.577$ is the Euler constant.
\footnote{In this
paper, for simplicity, we consider the case of a large ultraviolet
cutoff: $\Lambda^2\gg \bar\rho^2,~|eB|$, where $\bar\rho$ is a minimum
of the potential $V$.  Notice, however, that in some cases (as in
the application of the NJL model to hadron dynamics \cite{6}), $\Lambda^2$
and $\bar\rho^2$ may be comparable in magnitude.}

The gap equation $dV/d\rho=0$ is
\begin{eqnarray}
\rho\Lambda^2\biggl({1\over g}-1\biggr)&=&- \rho^3\ln
{(\Lambda\ell)^2\over 2} + \gamma\rho^3+\ell^{-2}\rho\ln
{(\rho\ell)^2\over 4\pi}\nonumber\\
&+&2\ell^{-2}\rho\ln \Gamma
\biggl({\rho^2\ell^2\over 2}\biggr) + O(1/\Lambda)~, \label{48}
\end{eqnarray}
\no where the dimensionless coupling constant $g=N_cG\Lambda^2/(4\pi^2)$.
In the derivation of this equation, we used the relations \cite{5}:
\begin{eqnarray}
{d\over dx} \zeta(\nu,x)=-\nu\zeta(\nu+1,x)~, \label{49}
\end{eqnarray}
\begin{eqnarray}
{d\over d\nu} \zeta(\nu,x)\bigl|_{\nu=0}=\ln \Gamma(x)
-{1\over 2}\ln 2\pi~, \quad
\zeta(0,x) = {1\over 2}-x~. \label{50}
\end{eqnarray}
\no As $B\to 0~(\ell\to\infty)$, we recover the known
gap equation in the NJL model \cite{3,6}:
\begin{eqnarray}
\rho\Lambda^2\biggl({1\over g}-1\biggr)
=-\rho^3 \ln{\Lambda\over \rho^2}~. \label{51}
\end{eqnarray}
\no This equation admits a nontrivial solution only if $g$ is
supercritical, $g>g_c=1$ (as Eq. (\ref{32}) implies, a solution to Eq.
(\ref{48}), $\rho=\bar\sigma$, coincides with the fermion dynamical mass,
$\bar\sigma=m_{\rm dyn}$, and the dispersion relation for fermions
is Eq. (\ref{11}) with $m$ replaced by $\bar\sigma$).  We will show that
the magntic field changes the situation dramatically: at $B\not= 0$,
a nontrivial solution exists for all $g>0$.

We shall first consider the case of subcritical $g$, $g<g_c=1$, which
in turn can be divided into two subcases: a) $g\ll g_c$ and 
b) $g\to g_c-0$ (nearcritical $g$).  Since at $g<g_c=1$, the
left-hand side in Eq. (\ref{48}) is positive and the first term on the
right-hand side in this equation is negative,  we conclude that a
nontrivial solution to this equation may exist only at
$\rho^2\ln(\Lambda\ell)^2 \ll \ell^{-2}\ln(\rho\ell)^{-2}$.  Then we
find the solution at $g\ll 1$:
\begin{eqnarray}
m^2_{\rm dyn} \equiv \bar\sigma^2={|eB|\over \pi} \exp
\biggl(-{4\pi^2(1-g)\over |eB|N_cG}\biggr)~. \label{52}
\end{eqnarray}
\no One can check that the dynamical mass is mostly generated in the
infrared region, with $k \lsim \ell^{-1}=|eB|^{1/2}$, where the
contribution of the LLL dominates.  In order to show this, one should
check that essentially the same result for $m_{\rm dyn}$ is obtained
if the full fermion propagator in the gap equation is replaced by the
LLL contribution and the cutoff $\Lambda$ is replaced by $\sqrt{|eB|}$
(see Section 8 below).

It is instructive to compare the relation (\ref{52}) with the relations for
the dynamical mass in the $(1+1)$--dimensional Gross--Neveu model \cite{13},
in the BCS theory of superconductivity \cite{4}, and in the 
$(2+1)$--dimensional NJL model in a magnetic field \cite{2}.

The relation for $m^2_{\rm dyn}$ in the Gross-Neveu model is
\begin{eqnarray}
m^2_{\rm dyn} = \Lambda^2\exp
\biggl(-{2\pi\over N_cG^{(0)}}\biggr) \label{53}
\end{eqnarray}
\no where $G^{(0)}$ is the bare coupling, which is dimensionless at
$D=1+1$. The similarity between relations (\ref{52}) and (\ref{53}) is evident:
$|eB|$ and $|eB|G$ in Eq. (\ref{52}) play the role of an ultraviolet cutoff
and the dimensionless coupling constant in Eq. (\ref{53}), respectively.
This of course reflects the point that the dynamics of the fermion
pairing in the $(3+1)$--dimensional NJL model in a magnetic field is
essentially $(1+1)$--dimensional. We shall return to the discussion of
the connection between this dynamics and that in the Gross-Neveu
model in Sec.~7.

We recall that, because of the Fermi surface, the dynamics of the 
electron in superconductivity is also $(1+1)$--dimensional.  This analogy
is rather deep.  In particular, the expression (\ref{52}) for $m_{\rm dyn}$
can be rewritten in a form similar to that for the energy gap
$\Delta$ in the
BCS theory: while 
$\Delta \sim \omega_D \exp\bigl(-{\rm const.}/\nu_FG_S\bigr)$, where
$\omega_D$ is the Debye frequency, $G_S$ is a coupling constant and
$\nu_F$ is the density of states on the Fermi surface, the mass
$m_{\rm dyn}$ is $m_{\rm dyn}\sim \sqrt{|eB|}~\exp\bigl(-1/2G\nu_0\bigr)$,
where the density of states $\nu_0$ on the energy surface $E=0$ of
the LLL is now given by expression (\ref{12}) multiplied by the factor $N_c$.
Thus the energy surface $E=0$ plays here the role of the Fermi surface.

Let us now compare the relation (\ref{52}) with that in the $(2+1)$--dimensional
NJL model in a magnetic field, in a weak coupling regime.  It is \cite{2}:
\begin{eqnarray}
m^2_{\rm dyn}=|eB|^2 {N_c^2G^2\over 4\pi^2} \label{54}
\end{eqnarray}
\no While the expression (\ref{52}) for $m^2_{\rm dyn}$ has an essential 
singularity at $G=0$, $m^2_{\rm dyn}$ in the $(2+1)$--dimensional NJL
model is analytic at $G=0$.  The latter is connected with the fact
that in 2+1 dimensions the condensate $\langle 0|\bar\psi\psi|0\rangle$
is non-zero even for free fermions in a magnetic field (see Eq.
(\ref{27})).
Indeed, Eq. (\ref{33}) implies that $m_{\rm dyn}=\langle 0|\sigma|0\rangle=
-G\langle0|\bar\psi\psi|0\rangle$.  From this fact, and Eq. (\ref{27}),
we get the relation (\ref{54}), to leading order in $G$.  Therefore the
dynamical mass $m_{\rm dyn}$ is essentially perturbative in $G$ in this
case.  As we will see in Section 8, this is in turn intimately
connected with the fact that, for D=2+1, the dynamics of fermion
pairing in a magntic field is $(0+1)$--dimensional.

Let us now consider near-critical values of $g$, with
$\Lambda^2(1-g)/g\rho^2\ll \ln(\Lambda\rho)^2$.  Looking for a
solution to the gap equation (\ref{48}) with $\rho^2\ell^2\ll 1$, we
arrive at the following equation:
\begin{eqnarray}
{1\over \rho^2\ell^2}\ln {1\over \pi\rho^2\ell^2}\simeq
\ln \Lambda^2\ell^2~, \label{55}
\end{eqnarray}
\no i.e.
\begin{eqnarray}
m^2_{\rm dyn}=\bar\sigma^2\simeq
|eB|{\ln\bigl[(\ln\Lambda^2\ell^2)/\pi\bigr]\over
\ln \Lambda^2\ell^2}~. \label{56}
\end{eqnarray}
\no What is the physical meaning of this relation?  Let us recall
that at $g=g_c$, the NJL model is equivalent to the (renormalizable)
Yukawa model \cite{6}.  In leading order in $1/N_c$, the renormalized Yukawa
coupling $\alpha_Y^{(\ell^{-1})}=(g_Y^{(\ell^{-1})})^2/(4\pi)$,
corresponding to the scale $\mu=\ell^{-1}$, is 
$\alpha_Y^{(\ell^{-1})}=\pi/\ln \Lambda^2\ell^2$ \cite{6}.  Therefore
the expression (\ref{56}) for the dynamical mass can be rewritten as
\begin{eqnarray}
m^2_{\rm dyn} \simeq |eB| {\alpha_Y^{(\ell^{-1})}\over \pi}
\ln {1\over \alpha_Y^{(\ell^{-1})}}~. \label{57}
\end{eqnarray}
\no Thus, as has to be the case in a renormalizable theory, the cutoff 
$\Lambda$ is removed, through the renormalization of parameters 
(the coupling constant, in this case), from the observable $m_{\rm dyn}$.

Let us now consider the case of supercritical values of $g$ 
($g>g_c$).  In this case an analytic expression for $m_{\rm dyn}$ can
be obtained for a weak magnetic field, satisfying the condition
$|eB|^{1/2}/m^{(0)}_{\rm dyn} \ll 1$, where $m^{(0)}_{\rm dyn}$ is
the solution to the gap equation (\ref{51}) with $B=0$.  Then we find from
Eq. (\ref{48}):
\begin{eqnarray}
m^2_{\rm dyn} \simeq (m^{(0)}_{\rm dyn})^2
\biggl[1+{|eB|^2\over 
3(m^{(0)}_{\rm dyn})^4 \ln(\Lambda/m^{(0)}_{\rm dyn})^2}\biggr]~,
\label{58}
\end{eqnarray}
\no i.e. $m_{\rm dyn}$ increases with $|B|$.
\footnote{The fact
that in the supercritical phase of the NJL model, $m_{\rm dyn}$ increases
with $|B|$ has been already pointed out by several authors (for a 
review, see Ref. \cite{15}).}  
Notice that in the near-critical region, with
$g-g_c\ll 1$, this expression for $m^2_{\rm dyn}$ can be rewritten as
\begin{eqnarray}
m^2_{\rm dyn} \simeq (m^{(0)}_{\rm dyn})^2
\biggl[1+{1\over 3\pi} \alpha_Y^{(m_{\rm dyn})}
{|eB|^2\over (m^{(0)}_{\rm dyn})^4}\biggr]~, \label{59}
\end{eqnarray}
where, to leading order in $1/N_c$, $\alpha_Y^{(m_{\rm dyn})}=
\pi/\ln(\Lambda/m_{\rm dyn})^2 \simeq \pi/\ln(\Lambda/m^{(0)}_{\rm dyn})^2$
is the renormalized Yukawa  coupling related to the scale
$\mu=m_{\rm dyn}$.

In conclusion, let us discuss the validity of the LLL dominance
approximation in more detail.  As Eq. (\ref{33}) implies, the dynamical
mass is
\begin{eqnarray}
m_{\rm dyn} = \langle 0|\sigma|0\rangle =
-G\langle 0|\bar\psi\psi|0\rangle~. \label{60}
\end{eqnarray}
\no Calculating the condensate $\langle 0|\bar\psi\psi|0\rangle$ in
the LLL dominance approximation, we find (see Eqs. (\ref{18}) and
(\ref{19})):
\begin{eqnarray}
m_{\rm dyn}=N_cG {m_{\rm dyn}|eB|\over 4\pi^2}
\ln {\Lambda^2\over m^2_{\rm dyn}}~, \label{61}
\end{eqnarray}
\no i.e.
\begin{eqnarray}
m^2_{\rm dyn}=\Lambda^2 \exp
\biggl(-{4\pi^2\over |eB|N_cG}\biggr)~. \label{62}
\end{eqnarray}
\no Comparing this relation with Eq. (\ref{52}), corresponding to dynamics
with $g$ outside the near--critical region, we conclude that this
approximation reproduces correctly the essential singularity in
$m^2_{\rm dyn}$ at $G=0$.  On the other hand, the coefficient of the
exponent in Eq. (\ref{52}) gets a contribution from all the Landau levels.

In the near-critical region, with $1-g \ll {m^2_{\rm dyn}\over \Lambda^2}
\ln \Lambda^2\ell^2$, all the Landau levels become equally important
(see Eq. (\ref{56})), and the LLL dominance approximation ceases to be valid.
This also happens (in a weak magnetic field)
at supercritical values of $g$, $g>g_c$, when the 
dynamical mass is generated even without a magnetic field.

Thus we conclude that, like the BCS approximation in the theory of
superconductivity, the LLL dominance approximation is appropriate for
the description of the dynamics of weakly interacting fermions.  This
point will be especially important in QED, considered in Section 9.
\ve

\ce {\bf 6. THE NJL MODEL IN A MAGNETIC FIELD:} 
\ce {\bf THE KINETIC TERM}
\b

Let us now consider the kinetic term ${\cal{L}}_k$ in the effective action
(\ref{34}).  The chiral $U_{\rm L}(1)\times U_{\rm R}(1)$ symmetry implies
that the general form of the kinetic term is
\begin{eqnarray}
{\cal{L}}_k = {F_1^{\mu\nu}\over 2} (\partial_\mu\rho_j \partial_\nu\rho_j)+
{F_2^{\mu\nu}\over \rho^2} (\rho_j\partial_\mu\rho_j)
(\rho_i\partial_\nu \rho_i)       \label{63}
\end{eqnarray}
\no where ${\mbox{\boldmath $\rho$}}=(\sigma,\pi)$ and
$F_1^{\mu\nu},F_2^{\mu\nu}$
are functions of $\rho^2$.  We found these functions by using the 
method of Ref. \cite{16}.  The derivation of ${\cal{L}}_k$ is considered in
Appendix A.  Here we shall present the final results.

The functions $F_1^{\mu\nu},~F_2^{\mu\nu}$ are $F_1^{\mu\nu}=
g^{\mu\nu}F_1^{\mu\mu},~F_2^{\mu\nu}=g^{\mu\nu} F_2^{\mu\mu}$ with
\begin{eqnarray}
F_1^{00}=F_1^{33}&=&{N_c\over 8\pi^2}
\biggl[\ln {\Lambda^2\ell^2\over 2}-\psi\biggl({\rho^2\ell^2\over 2}+1\biggr)
+{1\over \rho^2\ell^2}-\gamma+{1\over 3}\biggr]~,\nonumber\\ 
F_1^{11}=F_1^{22} &=& {N_c\over 8\pi^2}
\biggl[\ln {\Lambda^2\over \rho^2}-\gamma + {1\over 3}\biggr]~, \nonumber\\
F_2^{00}=F_2^{33}&=&-{N_c\over 24\pi^2}
\biggl[{\rho^2\ell^2\over 2} \zeta\biggl(2,~{\rho^2\ell^2\over 2}+1\biggr)
+{1\over \rho^2\ell^2}\biggr]~, \nonumber\\
F_2^{11}=F_2^{22} &=& {N_c\over 8\pi^2}
\biggl[\rho^4\ell^4\psi\biggl({\rho^2\ell^2\over 2}+1\biggr)
-2\rho^2\ell^2\ln \Gamma\biggl({\rho^2\ell^2\over 2}\biggr) \nonumber\\
&-& \rho^2\ell^2 \ln{\rho^2\ell^2\over 4\pi}
-\rho^4\ell^4-\rho^2\ell^2+1\biggr] \label{64}
\end{eqnarray}
\no where $\psi(x)=d(\ln\Gamma(x))/dx$.  (We recall that the magnetic
length is $\ell \equiv |eB|^{-1/2}$.)

As follows from Eqs. (\ref{63}) and (\ref{64}), the propagators of $\sigma$
and $\pi$ in leading order in $1/N_c$ have a genuine $(3+1)$--dimensional
form.  This agrees with the general arguments in support of the
$(3+1)$--dimensional form of propagators of neutral particles in a
magnetic field considered in Section 3.  We shall see in Appendix B
that this point is crucial for providing the reliability of the $1/N_c$
expansion in this problem.

Now, knowing the effective potential and the kinetic term, we can find
the dispersion law for the collective excitations $\sigma$ and $\pi$.
\ve

\ce {\bf 7. THE NJL MODEL IN A MAGNETIC FIELD: THE SPECTRUM}
\ce {\bf OF THE COLLECTIVE EXCITATIONS}
\b

The derivative expansion for the effective action is reliable in the
infrared region, with $k \lsim F_\pi$, where $F_\pi \sim m_{\rm dyn}$
is the decay constant of $\pi$.  Therefore it is appropriate for the
description of the low-energy dynamics of the gapless NG mode $\pi$,
but may, at most, aspire to describe only qualitatively the dynamics
of the $\sigma$ mode.  Nevertheless, for completeness, we will derive
the dispersion law both for $\pi$ and $\sigma$.

We begin by considering the spectrum of the collective excitations
in the subcritical, $g<g_c$, region.  At $g \ll g_c=1$ we find the following
dispersion laws from Eqs. (\ref{47}), (\ref{63}) and (\ref{64}):
\begin{eqnarray}
E_\pi & \simeq& \biggl[\frac{m_{\rm dyn}^2}{|eB|}
\ln\biggl(\frac{|eB|}{\pi m_{\rm dyn}^2}\biggr)
{\bf k}_{\perp}^2+k_3^2\biggr]^{1/2}~.   \label{65}\\
E_{\sigma} & \simeq& \biggl[12~m^2_{\rm dyn} +
\frac{3m_{\rm dyn}^2}{|eB|}\ln\biggl(\frac{|eB|}{\pi m_{\rm
dyn}^2}\biggr)
{\bf k}^2_{\perp}+k_3^2)\biggr]^{1/2}  \label{66}  
\end{eqnarray}
\no with $m_{\rm dyn}$ defined in Eq. (\ref{52}).  Thus the $\pi$ is a gapless
NG mode.  Taking into account Eq. (\ref{52}), we find that the transverse
velocity $|{\bf v}_{\perp}|=|\partial E_\pi/\partial {\bf k}_{\perp}|$
of $\pi$ is less than 1.

In the near--critical region, with $g-g_c \ll 1$ (where the NJL model
is equivalent to the Yukawa model), we find from Eqs. (\ref{47}),
(\ref{57}), (\ref{63}) and (\ref{64}):
\begin{eqnarray}
E_\pi & \simeq & \biggl[\biggl(1-
{1\over \ln \pi/\alpha_Y^{(\ell^{-1})}}\biggr)  
{\bf k}^2_{\perp}+k_3^2\biggr]^{1/2}~, \label{67}\\
E_\sigma &=& \biggl[4m^2_{\rm dyn}
\biggl(1+{2\over 3\ln \pi/\alpha_Y^{(\ell^{-1})}}\biggr)+
\biggl(1-{1\over 3\ln \pi/\alpha_Y^{(\ell^{-1})}}\biggr)
{\bf k}^2_{\perp}+k_3^2\biggr]^{1/2}~.    \label{68}
\end{eqnarray}
\no Thus, as has to be the case in a renormalizable model, the cutoff 
$\Lambda$ disappears from the observables $E_\pi$ and $E_\sigma$.  
Note also that the transverse velocity $v_{\perp}$ of $\pi$ is again 
less than 1.

Since the derivative expansion is valid only at small values of 
$k_{\perp},k_3$, satisfying \break 
$k_{\perp},k_3 \lsim F_\pi \sim m_{\rm dyn}$,
the relations (\ref{65})-(\ref{68}) are valid only in the infrared region (as we 
already indicated above, the relations (\ref{66}) and (\ref{68}) for $\sigma$ are 
only estimates).  In particular, since, as $B\to 0$, the
dynamical mass $m_{\rm dyn}$ and the decay constant $F_\pi \sim m_{\rm dyn}$
go to zero in the subcritical region, these relations cease to be valid,
even in the infrared region, at $B=0$.

Let us now turn to the supercritical region, with $g>g_c$.  The analytic
expressions for $E_\pi$ and $E_\sigma$ can be obtained for weak magnetic
fields, satisfying $|eB|^{1/2} \ll m^{(0)}_{\rm dyn}$, where
$m^{(0)}_{\rm dyn}$ is the dynamical mass of fermions at $B=0$.  We
find from Eqs. (\ref{47}), (\ref{63}) and (\ref{64}):
\begin{eqnarray}
E_\pi &\simeq& \biggl[\biggl(1-
{(eB)^2\over 3m^4_{\rm dyn} \ln(\Lambda/m_{\rm dyn})^2}\biggr)
{\bf k}^2_{\perp}+k_3^2\biggr]^{1/2}~,\label{69} \\
E_\sigma &\simeq&\biggl[4m^2_{\rm dyn}\biggl(1+
{1\over 3\ln(\Lambda/m_{\rm dyn})^2} +
{4(eB)^2\over 9m^4_{\rm dyn} \ln(\Lambda/m_{\rm dyn})^2}\biggr) 
\nonumber\\
&+&\biggl(1-{11(eB)^2\over 45m^4_{\rm dyn}\ln(\Lambda/m_{\rm dyn})^2}\biggr)
{\bf k}^2_{\perp}+k^2_3\biggr]^{1/2}~,   \label{70}
\end{eqnarray}
\no with $m_{\rm dyn}$ given in Eq. (\ref{58}).  One can see that the
magnetic field reduces the transverse velocity $v_{\perp}$ of $\pi$
in this case as well.

Since in the supercritical region, with $g-g_c \ll 1$, the renormalized 
Yukawa coupling is $\alpha_Y^{(m_{\rm dyn})}=\pi/\ln(\Lambda/m_{\rm
dyn})^2$ to leading
order in $1/N_c$, Eqs. (\ref{69}) and (\ref{70}) can be rewritten in that region
as
\begin{eqnarray}
E_\pi &\simeq &\biggl[\biggl(1-\alpha_Y^{(m_{\rm dyn})}
{(eB)^2\over 3\pi m^4_{\rm dyn}}\biggr)
{\bf k}^2_{\perp}+k_3^2\biggr]^{1/2}~, \label{71}\\
E_\sigma &\simeq &\biggl[4m^2_{\rm dyn}\biggl(1+\alpha_Y^{(m_{\rm dyn})}
\biggl({1\over 3\pi}+{4(eB)^2\over 9\pi m^4_{\rm dyn}}\biggr)\biggr) 
\nonumber\\
&+&\biggl(1-\alpha_Y^{(m_{\rm dyn})}
{11(eB)^2\over 45\pi m^4_{\rm dyn}}\biggr) 
{\bf k}^2_{\perp}+k_3^2\biggr]^{1/2}~.    \label{72}
\end{eqnarray}

Let us now return to the discussion of the connection between the
dynamics in the $(3+1)$--dimensional NJL model in a magnetic field
(with a weak coupling $g\ll 1$) and that in the Gross--Neveu (GN)
model. Comparison of Eqs.(\ref{52}) and (\ref{53}) suggests that the infrared
dynamics in these two models may be similar. The GN model is
asymptotically free, with the bare coupling constant
$G^{(0)}=2\pi/N_c\ln(\Lambda^2/m^2_{\rm dyn})\to 0$ as $\Lambda\to
\infty$. Therefore there is the dimensional transmutation in the
model: in the scaling region, at $G^{(0)}\ll 1$, the infrared dynamics with 
$(\ln\Lambda^2/k^2)^{-1}\ll 1$ is essentially independent either of
the coupling constant $G^{(0)}$ or the cutoff $\Lambda$; the only
relevant parameter is the dynamical mass $m_{\rm dyn}$ (which is an
analogue of the parameter $\Lambda_{QCD}$ in QCD). One might expect
that a similar dimensional transmutation should take place in the
$(3+1)$--dimensional NJL model in a magnetic field: at 
$|eB|GN_c\ll 1$ 
(see Eq.(\ref{52})), the infrared dynamics should be essentially independent 
of the magnetic field and the coupling constant $G$.

The situation in this model is however more subtle. As follows from
Eq.(\ref{18}) (with $m$ replaced by $m_{\rm dyn}$), the fermion propagator
in the infrared region is indeed independent (up to power corrections 
$\sim ({\bf k}^2_{\perp}/|eB|)^n$, $n\geq 1$) of the magnetic field.
However, the dependence of the propagator of the NG mode $\pi$ on
$|eB|$ is essential. Though, as follows from Eqs.(\ref{52}) and
(\ref{65}), the maximum transverse velocity 
$|{\bf v}_{\perp}|=\sqrt{(m_{\rm dyn}^2/|eB|)\ln(|eB|/\pi m_{\rm dyn}^2)}$
of $\pi$ is small at $|eB|GN_c\ll 1$, it is crucial that it is nonzero. 
The latter
provides the $(3+1)$--dimensional character of the NG propagator and,
as was already explained in Sec.~3, this in turn is crucial for the
realization of spontaneous chiral symmetry breaking in the model.

We recall that in the $(1+1)$--dimensional GN model with the chiral
$U_{L}(1)\times U_{R}(1)$ symmetry, there is no spontaneous chiral
symmetry breaking, despite the fact that 
the fermion dynamical mass is nonzero:
in accordance with the MWC theorem \cite{12}, there are no NG bosons in
this model. As was suggested by Witten (see the second paper in
Ref.\cite{13}), in the GN model, the Berezinsky-Kosterlitz-Thouless (BKT)
phase is presumably realized, with the NG bosons being replaced by
the BKT vortex collective excitations.

Therefore, in $(3+1)$--dimensional NJL model, the magnetic length
$\ell\equiv |eB|^{-1/2}$ acts as a (physical) regulator: at finite $\ell$,
there is spontaneous chiral symmetry breaking, the propagator of NG
mode $\pi$ has a $(3+1)$--dimensional form and, as a result, the
$1/N_c$ expansion is reliable (see Appendix B). In a sense,
the magnetic length $\ell$  plays here the same role as the
$\epsilon$ parameter in the $(2+\epsilon)$--dimensional GN model 
\cite{13}\footnote{Comparing the effective actions in the
$(3+1)$--dimentsional NJL model in a magnetic field and 
$(1+1)$--dimentsional GN model, one can show (at least formally)
that, as $|eB|\to \infty$, the NJL model, with $N_c$ colors becomes
equivalent to the GN model with the number of colors
$\tilde{N}_c=N_cN$, where $N=|eB|S_{12}/2\pi\to \infty$ is the number
of states at the LLL ($S_{12}$ is the square in the $x_1x_2$-plane).
This interesting limit will be considered in more detail elsewhere.}.

It is noticeable that the observables $E_{\pi}$ and $E_{\sigma}$ in
Eqs.(\ref{65}) and (\ref{66}) do not depend explicitly on cutoff
$\Lambda$. It happens because, due to the relation (\ref{52}) for
$m_{\rm dyn}$, the term 
$1/\rho^2\ell^2|_{\rho^2\simeq\bar{\rho}^2=m^2_{\rm dyn}}$ dominates
in the functions $F^{00}_{1}=F^{33}_{1}$ and $F^{00}_{2}=F^{33}_{2}$
in the kinetic term (see Eq.(\ref{64})). In this dynamical regime,
there is a strong hierachy of two scales: $m_{\rm dyn}^2\ll
|eB|$. In a sense, with respect to the infrared dynamics with
$k^2\ll |eB|$, the scale $|eB|$ plays the role of a (physical)
ultraviolet cutoff.

Let us consider the renormalization group properties of this
dynamics.

The reliability of the $1/N_c$ expansion implies that the infrared dynamics
in this model is (at least at sufficiently
large $N_c$) under control: the effects connected with nonleading
orders in $1/N_c$ are small (see Appendix B). Let us consider the
running Yukawa coupling in this model. In leading order
in  $1/N_c$, the bare Yukawa coupling and both the renormalization 
constant of the fermion field
and that of the Yukawa vertex equal one (see Appendix B). Therefore
the behavior of the running Yukawa coupling is determined by the
renormalization constant of the chiral field $\mbox{\boldmath$\rho$}$: 
$g^{(\mu)}_Y\sim \sqrt{Z^{(\mu)}_{\rho}}$. Eq.(\ref{64}) implies that 
$Z^{(\mu)^{-1}}_{\rho}\sim N_c|eB|/8\pi^2\mu^2$ at $\mu\sim m_{\rm
dyn}$. Therefore the running Yukawa coupling is weak in the infrared
region: $g^{(\mu)}_{Y}\sim \sqrt{8\pi^2\mu^2/N_c|eB|}$ and 
$g^{(\mu)}_{Y}\to 0$ as $|eB|$ (and therefore also $\Lambda$) goes to infinity, 
{\em i.e.} there is a Gaussian infrared fixed point.

Examining the effective potential and the kinetic term in the
effective action, one can show that a similar behavior occurs for the
rest of running couplings, describing self--interactions of
the chiral field $\mbox{\boldmath$\rho$}$.

We emphasize that this discussion relates only to the NJL model with
a weak coupling constant. In the case of the NJL model 
with a near--critical $g$, the situation is different. In that
case, one finds from (\ref{64}) that the renormalization constant 
$Z^{(\mu)^{-1}}_{\rho}$ is 
$Z^{(\mu)^{-1}}_{\rho}\sim (N_c/4\pi^2)\ln(\Lambda/\mu)$. Therefore
the behavior of the running coupling is similar to that in the NJL
model without magnetic field (and with the same near--critical $g$)
\cite{6}: 
$g^{(\mu)}_{Y}\sim 2\pi/\sqrt{N_c\ln\Lambda/\mu}$, {\em
i.e.} there is the usual Gaussian infrared fixed point in this case.

As was already indicated in the previous section, 
the difference between these two
dynamical regimes is connected with the point that, while at $g\ll 1$
the LLL dominates, at a near--critical $g$, all Landau levels become
relevant.

In the next section, we will derive the Bethe--Salpeter equation for
the NG mode $\pi$ in the LLL
dominance approximation.  This will yield a further insight into the
physics of the dimensional reduction $D\to D-2$ in the dynamics
of fermion pairing in a magnetic field.
\ve

\ce {\bf 8. THE NJL MODEL IN A MAGNETIC FIELD:} 
\ce {\bf THE BETHE-SALPETER EQUATION}
\ce {\bf FOR THE NG MODE}
\b

The homogeneous Bethe-Salpeter (BS) equation for the NG bound state
$\pi$ takes the form (for a review, see Ref. \cite{6}):
\begin{eqnarray}
\chi_{AB}(x,y;P) &=& -i\int d^4x_1d^4y_1d^4x_2d^4y_2G_{AA_1}(x,x_1)
K_{A_1B_1;A_2B_2}(x_1y_1,x_2y_2) \nonumber\\
&\times& \chi_{A_2B_2}(x_2,y_2;P)G_{B_1B}(y_1,y)~, \label{74}
\end{eqnarray}

\no where the BS wave function $\chi_{AB}(x,y;P)=
\langle 0|T\psi_A(x)\bar\psi_B(y)|P;\pi\rangle$ and the fer\-mi\-on
propagator $G_{AB}(x,y)=\langle 0|T\psi_A(x)\bar\psi_B(y)|0\rangle$;
the indices $A=(n\alpha)$ and $B=(m\beta)$ include both Dirac $(n,m)$
and color $(\alpha,\beta)$ indices.  Notice that though the external
field $A_\mu^{\rm ext}$ breaks the conventional translation invariance,
the total momentum $P$ is a good, conserved, quantum number for neutral
bound states, in particular for the $\pi$ (see Section 3).  Henceforth
we will use the symmetric gauge (\ref{10}).

In leading order in $1/N_c$, the BS kernel in the NJL model is \cite{6}:
\begin{eqnarray}
K_{A_1B_1;A_2B_2}(x_1y_1,x_2y_2) &=& G\bigl[
\delta_{A_1B_1}\delta_{B_2A_2}+\delta_{\alpha_1\beta_1}
\delta_{\beta_2\alpha_2}(i\gamma_5)_{n_1m_1}
(i\gamma_5)_{m_2n_2} \nonumber\\
&-& \delta_{A_1A_2}\delta_{B_2B_1}-\delta_{\alpha_1\alpha_2}
\delta_{\beta_2\beta_1}(i\gamma_5)_{n_1n_2}
(i\gamma_5)_{m_2m_1}\bigr] \nonumber\\
&\times& \delta^4(x_1-y_1)\delta^4(x_1-x_2)
\delta^4(x_1-y_2) ~. \label{75}
\end{eqnarray}

\no Also, as was already indicated in Section 5, in this approximation,
the full-fermion propagator coincides with the propagator $S$
(\ref{13}) of
a free fermion (with $m=m_{\rm dyn}$) in a magnetic field.

Let us now factorize (as in Eq. (\ref{13})) the Schwinger phase factor in
the BS wave function:
\begin{eqnarray}
\chi_{AB}(R,r;P) = \delta_{\alpha\beta} \exp
(ier^\mu A_\mu^{\rm ext}(R)) e^{-iPR} \tilde\chi_{nm}(R,r;P)~,
\label{76}
\end{eqnarray}
\no where we introduced the relative coordinate, $r=x-y$, and the center
of mass coordinate, $R=(x+y)/2$.  Then Eq. (\ref{74}) can be rewritten as
\begin{eqnarray}
\tilde\chi_{nm}(R,r;P)&=&-iN_cG \int d^4R_1
\tilde S_{nn_1}\biggl({r\over 2}+R-R_1\biggr)
\biggl[\delta_{n_1m_1}~{\rm tr}(\tilde\chi(R_1,0;P)) \nonumber\\
&-&(\gamma_5)_{n_1m_1}{\rm tr}(\gamma_5\tilde\chi(R_1,0;P))
-{1\over N_c}\tilde\chi_{n_1m_1}(R_1,0;P)  \nonumber\\
&+&{1\over N_c}
(\gamma_5)_{n_1n_2}\tilde\chi_{n_2m_2}(R_1,0;P) 
(\gamma_5)_{m_2m_1}\biggr] \nonumber\\
&&\hspace{-20mm}\times \tilde S_{m_1m}\biggl({r\over 2}-R+R_1\biggr)
\exp\bigl[-ier^\mu A_\mu^{\rm ext}(R-R_1)\bigr]
\exp\bigl[iP(R-R_1)\bigr]~.   \label{77}
\end{eqnarray}
\no The important fact is that the effect of translation symmetry
breaking by the magnetic field is factorized in the Schwinger phase
factor in Eq. (\ref{76}), and Eq. (\ref{77}) admits a translation invariant
solution: $\tilde\chi_{nm}(R,r;P)=\tilde\chi_{nm}(r,P)$.
\footnote
{\no This point is intimately connected with the fact that these bound
states are neutral: the Schwinger phase factor is universal for
neutral fermion-antifermion bound states.}
Then, transforming this equation into momentum space, we get:
\begin{eqnarray}
\tilde\chi_{nm}(p;P) &=&-iN_cG \int
{d^2q_{\perp}d^2R_{\perp}d^2k_{\perp}d^2k_{\parallel}\over (2\pi)^6}
\nonumber\\
&\times& \exp\bigl[i({\bf P}_{\perp}-{\bf q}_{\perp})
{\bf R}_{\perp}\bigr] \tilde S_{nn_1}
\biggl(p_{\parallel}+{P_{\parallel}\over 2},{\bf p}_{\perp}
+e {\bf A}^{\rm ext}({\bf R}_{\perp}) +
{{\bf q}_{\perp}\over 2}\biggr) \nonumber\\
&\times& \biggl[\delta_{n_1m_1}~{\rm tr}\bigl(\tilde\chi(k;P)\bigr)-
(\gamma_5)_{n_1m_1}~{\rm tr}\bigl(\gamma_5\tilde\chi(x;P)\bigr)-
{1\over N_c} \tilde\chi_{n_1m_1}(k;P) \nonumber\\
&+& {1\over N_c}(\gamma_5)_{n_1n_2}\tilde\chi_{n_2m_2}(k;P)
(\gamma_5)_{m_2m_1}\biggr] \nonumber\\
&\times &\tilde S_{m_1m}\biggl(p_{\parallel}-{P_{\parallel}\over 2},
{\bf p}_{\perp}+
e{\bf A}^{\rm ext}({\bf R}_{\perp})-{{\bf q}_{\perp}\over 2}\biggr)~,
\label{78}
\end{eqnarray}

\no where $p_{\parallel} \equiv (p^0,p^3),~{\bf p}_{\perp} \equiv
(p^1,p^2)$.  Henceforth we will consider the equation with the total
momentum $P_\mu \to 0$.

We shall consider the case of weakly interacting fermions, when the
LLL pole approximation for the fermion propagator is justified.
Henceforth, for concreteness, we will consider the case $eB>0$. Then
\begin{eqnarray}\tilde S(p) \simeq i \exp(-\ell^2 {\bf p}^2_{\perp})
{\hat p_{\parallel}+m_{\rm dyn}\over p_{\parallel}^2-
m^2_{\rm dyn}}(1-i\gamma^1\gamma^2) \label{79}
\end{eqnarray}
\no (see Eq. (\ref{18})), where $\hat p_{\parallel}=p^0\gamma^0-p^3\gamma^3$,
and Eq. (\ref{78}) transforms into the following one:
\begin{eqnarray}
\rho(p_{\parallel},{\bf p}_{\perp}) &=& {iN_cG\ell^2\over (2\pi)^5}
e^{-\ell^2{\bf p}^2_{\perp}} \int d^2A_{\perp}d^2k_{\perp}d^2k_{\parallel}
e^{-\ell^2{\bf A}^2_{\perp}} (1-i\gamma^1\gamma^2) \nonumber\\
&\times& \hat F\bigl[\rho(k_{\parallel},{\bf k}_{\perp})\bigr]
(1-i\gamma^1\gamma^2)~, \label{80}
\end{eqnarray}
\no where
\begin{eqnarray}
\rho(p_{\parallel},{\bf p}_{\perp})=
(\hat p_{\parallel}-m_{\rm dyn})\tilde\chi(p_{\parallel},
{\bf p}_{\perp})(\hat p_{\parallel}-m_{\rm dyn}) \label{81}
\end{eqnarray}
\no with $\chi(p_{\parallel},{\bf p}_{\perp}) \equiv \chi
(p_{\parallel},{\bf p}_{\perp};P)\bigl|_{P=0}$,
and the operator symbol $\hat F[\rho]$ means:
\begin{eqnarray}
\hat F\bigl[\rho(k_{\parallel},{\bf k}_{\perp})\bigr]&=&
{\rm tr}\biggl({\hat k_{\parallel}+m_{\rm dyn}\over
k^2_{\parallel}-m^2_{\rm dyn}}\rho(k_{\parallel},{\bf k}_{\perp})
{\hat k_{\parallel}+m_{\rm dyn}\over k^2_{\parallel}-m^2_{\rm dyn}}\biggr)
\nonumber\\
&-&\gamma_5 {\rm tr}\biggl(\gamma_5{\hat k_{\parallel}+m_{\rm dyn}\over
k^2_{\parallel}-m_{\rm dyn}} \rho(k_{\parallel},{\bf k}_{\perp})
{\hat k_{\parallel}+m_{\rm dyn}\over k^2_{\parallel}-m^2_{\rm dyn}}\biggr) 
\nonumber\\
&-&{1\over N_c} {\hat k_{\parallel}+m_{\rm dyn}\over k^2_{\parallel}-
m^2_{\rm dyn}}
\rho(k_{\parallel},{\bf k}_{\perp})
{\hat k_{\parallel}+m_{\rm dyn}\over k^2_{\parallel}-m^2_{\rm dyn}} 
\nonumber\\
&+&{1\over N_c} \gamma_5 {\hat k_{\parallel}+m_{\rm dyn}\over k^2_{\parallel}
-m^2_{\rm dyn}}\rho(k_{\parallel},{\bf k}_{\perp})
{\hat k_{\parallel}+m_{\rm dyn}\over k^2_{\parallel}-m^2_{\rm dyn}}
\gamma_5~. 
\end{eqnarray}

Eq. (\ref{80}) implies that $\rho(p_{\parallel},{\bf p}_{\perp})=
\exp(-\ell^2 {\bf p}^2_{\perp})\varphi(p_{\parallel})$, where
$\varphi(p_{\parallel})$ satisfies the equation:
\begin{eqnarray}
\varphi(p_{\parallel})={iN_cG\over 32\pi^3\ell^2} \int
d^2k_{\parallel}(1-i\gamma^1\gamma^2)
\hat F\bigl[\varphi(k_{\parallel})\bigr]
(1-i\gamma^1\gamma^2)~. \label{83}
\end{eqnarray}
\no Thus the BS equation has been reduced to a two--dimensional 
integral equation.  Of course, this fact reflects the two--dimensional
character of the dynamics of the LLL, that can be explicitly read off
from Eq. (\ref{79}).

Henceforth we will use Euclidean space with $k_4=-ik^0$. 
In order to define
the matrix structure of the wavefunction $\varphi(p_{\parallel})$ of
the $\pi$, note that in a magnetic field, in the symmetric gauge
(\ref{10}),
there is the symmetry $SO(2)\times SO(2)\times {\cal{P}}$, where the
$SO(2)\times SO(2)$ is connected with rotations in the $x_1-x_2$
and $x_3-x_4$ planes and ${\cal{P}}$ is the inversion transformation
$x_3\to -x_3$ (under which a fermion field transforms as
$\psi\to i\gamma_5 \gamma_3\psi)$.  This symmetry implies that
the function $\varphi(p_{\parallel})$ takes the form:
\begin{eqnarray}
\varphi(p_{\parallel})=\gamma_5(A+i\gamma_1\gamma_2
B+\hat p_{\parallel}C+i\gamma_1\gamma_2 \hat p_{\parallel} D)
\label{84}
\end{eqnarray}
\no where $\hat p_{\parallel}=p_3\gamma_3+p_4\gamma_4$ and $A,B,C$ 
and $D$ are functions of $p^2_{\parallel}$ ($\gamma_\mu$ are
antihermitian in Euclidean space).  Substituting expansion (\ref{84})
into Eq. (\ref{83}), we find that $B=-A$, $C=D=0$, i.e.
$\varphi(p_{\parallel})=A\gamma_5(1-i\gamma_1\gamma_2)$.
\footnote
{The occurrence of the projection operator $(1-i\gamma^1\gamma^2)/2$
in the wave function $\varphi$ reflects the fact that the spin of
fermions at the LLL is polarized along the magnetic field.\label{spin}}
The function $A$ satisfies the equation
\begin{eqnarray}
A(p)={N_cG\over 4\pi^3\ell^2} \int d^2k
{A(k)\over k^2+m^2_{\rm dyn}}~. \label{85}
\end{eqnarray}
\no The solution to this equation is $A(p)={\rm constant}$, and 
introducing the ultraviolet cutoff $\Lambda$, we get the gap
equation for $m^2_{\rm dyn}$:
\begin{eqnarray}
1={N_cG\over 4\pi^2\ell^2} \int^{\Lambda^2}_0
{dk^2\over k^2+m^2_{\rm dyn}}~. \label{86}
\end{eqnarray}                         
\no It leads to the expression
\begin{eqnarray*}
m^2_{\rm dyn}=\Lambda^2\exp\biggl(-{4\pi^2\ell^2\over N_cG}\biggr)
\end{eqnarray*}
\no for $m^2_{\rm dyn}$ which coincides with expression (\ref{62}) derived
in Section 5.

The integral equation (\ref{85}) can be rewritten in the form of a
two--dimensional \break Schr\"odinger equation with a $\delta$--like potential.
Indeed, introducing the wave function
\begin{eqnarray}
\Psi({\bf r})=\int {d^2k\over (2\pi)^2}
{e^{-i{\bf k r}}\over k^2+m^2_{\rm dyn}} A(k)~, \label{87}
\end{eqnarray}
\no we find
\begin{eqnarray}
\biggl(-\Delta+m^2_{\rm dyn}-{N_cG\over \pi\ell^2}
\delta^2_\Lambda({\bf r})\biggr)\Psi({\bf r})=0 \label{88}
\end{eqnarray}
\no where
\begin{eqnarray*}
\Delta={\partial^2\over \partial r^2_1}+
{\partial^2\over \partial r^2_2} 
\end{eqnarray*}
\no and
\begin{eqnarray}
\delta^2_\Lambda({\bf r})=\int_\Lambda
{d^2k\over (2\pi)^2} e^{-i{\bf kr}} \label{89}
\end{eqnarray}
\no Notice that in the same way, one can show that in the 
$(2+1)$--dimensional NJL model \cite{2}, the analog of Eq. (\ref{88}) has the
form of a one--dimensional Schr\"odinger equation:
\begin{eqnarray}
\biggl(-{\partial^2\over \partial r^2}+m^2_{\rm dyn}-
{N_cG\over \pi^2} \delta_\Lambda(r)\biggr)\Psi(r)=0 \label{90}
\end{eqnarray}
\no with $\delta_\Lambda(r)=\int^\Lambda_{-\Lambda}
{dk\over 2\pi} e^{-ikr}$.  It leads to the following gap equation
for $m^2_{\rm dyn}$:
\begin{eqnarray}
1={N_cG\over 2\pi^2\ell^2} \int^\Lambda_{-\Lambda}
{dk\over k^2+m^2_{\rm dyn}}~, \label{91}
\end{eqnarray}
\no which yields expression (\ref{4}) for $m^2_{\rm dyn}$ in the limit
$\Lambda\to\infty$: $m^2_{\rm dyn}=(|eB|N_cG/2\pi)^2$.

Thus, since $(-m^2_{\rm dyn})$ plays the role of the energy $E$ in the
Schr\"odinger equation with a negative, i.e. attractive, potential,
the problem has been reduced to finding the spectrum of bound
states (with $E=-m^2_{\rm dyn}<0$) of the two--dimensional and
one--dimensional Schr\"odinger equation with a short--range,
$\delta$-like, potential.  (Actually, since only the largest value
of $m^2_{\rm dyn}$ defines the stable vacuum \cite{6}, one has to find
the value of the energy for the lowest stationary level.)  This allows
us to use some general results proved in the literature that will
in turn yield some additional insight into the dynamics of chiral 
symmetry breaking by a magnetic field.

First, there is at least one bound state for the one--dimensional
($d=1$) and two--dimensional ($d=2$) Schr\"odinger equation with an
attractive potential \cite{7}.  Second, while the energy of the lowest
level $E(G)$ is an analytic function of the coupling constant, around
$G=0$, at $d=1$, it is non--analytic at $G=0$ at $d=2$ \cite{7}.  Moreover,
at $d=2$ for short--range potentials, the energy $E(G)=-m^2_{\rm dyn}(G)$
takes the form $E(G)\sim -\exp\bigl[1/(aG)\bigr]$ (with $a$ being a
positive constant) as $G\to 0$ \cite{7}.

Thus the results obtained in the NJL model at $D=2+1$ and $D=3+1$ agree
with these general results.

The fact that the effect of spontaneous chiral symmetry breaking by a
magnetic field is based on the dimensional reduction $D\to D-2$
in a magnetic field suggests that this effect is general, and not
restricted to the NJL model.  In the next section, we shall consider
this effect in $(3+1)$--dimensional QED.
\ve

\ce {\bf 9. SPONTANEOUS CHIRAL SYMMETRY BREAKING}
\ce {\bf BY A MAGNETIC FIELD IN QED}
\b

The dynamics of fermions in a constant magnetic field in QED was first
considered by Schwinger \cite{1}.  In that classical work, while
the interaction with the external magnetic field was considered in
all orders in the coupling cosntant, the quantum dynamics was treated
perturbatively.  There is no dynamical chiral symmetry breaking in
this approximation \cite{17}.  In this section we reconsider this problem,
treating the QED dynamics non-perturbatively, and show that, in ladder 
and improved ladder approximations, a constant magnetic field leads 
to spontaneous chiral symmetry breaking.

We will use the same strategy as in the previous section and derive the
BS equation for the NG modes.

The Lagrangian density of massless QED in a magnetic field is:
\begin{eqnarray}
{\cal{L}}=-{1\over 4}F^{\mu\nu}F_{\mu\nu}+{1\over 2}\bigl[\bar\psi,
(i\gamma^\mu D_\mu)\psi\bigr]~, \label{92}
\end{eqnarray}
\no where the covariant derivative $D_\mu$ is
\begin{eqnarray}
D_\mu=\partial_\mu-ie(A_\mu^{\rm ext}+A_\mu)~, \quad
A_\mu^{\rm ext}=\biggl(0,~-{B\over 2} x_2,~{B\over 2} x_1,~0\biggr)~,
\label{93}
\end{eqnarray}
\no i.e. we use the symmetric gauge (\ref{10}).  Besides the Dirac index 
($n$), the fermion field carries an additional flavor
index $a=1,2,\ldots,N_f$.  Then the Lagrangian density (\ref{92}) is
invariant under the chiral $SU_{\rm L}(N_f)\times SU_{\rm R}(N_f)\times
U_{\rm L+R}(1)$ (we will not discuss the dynamics related to the
anomalous, singlet, axial current  $J_{5\mu}$ in this paper).
Since we consider the weak coupling phase of QED, there is no 
spontaneous chiral symmetry breaking at $B=0$ \cite{6,18}.  We will show
that the magnetic field changes the situation dramatically: at $B\not= 0$
the chiral symmetry $SU_{\rm L}(N_f)\times SU_{\rm R}(N_f)$ breaks down to
$SU_{\rm V}(N_f) \equiv SU_{\rm L+R}(N_f)$.  As a result, the dynamical
mass $m_{\rm dyn}$ is generated, and $N_f^2-1$ gapless
bosons, composed of fermions and antifermions, appear.

The homogeneous BS equation for the $N_f^2-1$ NG bound states takes the
form \cite{6}:
\begin{eqnarray}
\chi^\beta_{AB}(x,y;P) &=& -i \int d^4x_1 d^4y_1
d^4x_2d^4y_2 G_{AA_1}(x,x_1) 
K_{A_1B_1;A_2B_2}(x_1y_1,x_2y_2) \nonumber\\
&\times& \chi^\beta_{A_2B_2}(x_2,y_2;P)
G_{B_1B}(y_2,y)~,    \label{94}
\end{eqnarray}

\no where the BS wave function $\chi^\beta_{AB}=
\langle0|T\psi_A(x)\bar\psi_B(y)|P;\beta\rangle$, $\beta=1,\ldots,
N_f^2-1$, and the fermion propagator $G_{AB}(x,y)=
\langle0|T\psi_A(x)\bar\psi_B(y)|0\rangle$; the indices
$A=(na)$ and $B=(mb)$ include both Dirac $(n,m)$ and flavor
$(a,b)$ indices.

As will be shown below, the NG bosons are formed in the
infrared region, where the QED coupling is weak. This seems suggest that
the BS kernel in leading order in $\alpha$ is a reliable approximation. 
However, because of the $(1+1)$--dimensional form of the fermion 
propagator in the infrared region, there may be also relevant 
higher order contributions. We shall return to this problem below,
but first, we shall analyze the BS equation with the kernel in leading 
order in $\alpha$.

The BS kernel in leading order in $\alpha$ is \cite{6}:
\begin{eqnarray}
K_{A_1B_1;A_2B_2}(x_1y_1,x_2,y_2)&=&
-4\pi i\alpha \delta_{a_1a_2}\delta_{b_2b_1}\gamma^\mu_{n_1n_2}
\gamma^\nu_{m_2m_1}{\cal{D}}_{\mu\nu}(y_2-x_2) \nonumber\\
&&\hspace{-20mm}\times\delta(x_1-x_2)\delta(y_1-y_2)+4\pi 
i\alpha\delta_{a_1b_1}
\delta_{b_2a_2}\gamma^\mu_{n_1m_1}\gamma^\nu_{m_2n_2}
{\cal{D}}_{\mu\nu}(x_1-x_2) \nonumber\\
&&\hspace{-20mm}\times \delta(x_1-y_1)\delta(x_2-y_2)~,   \label{95}
\end{eqnarray}
\no where the photon propagator
\begin{eqnarray}
{\cal{D}}_{\mu\nu}(x)={-i\over (2\pi)^4} \int d^4k e^{ikx}
\biggl(g_{\mu\nu}-\lambda {k_\mu k_\nu \over k^2}\biggr)
{1\over k^2}                         \label{96}
\end{eqnarray}
\no ($\lambda$ is a gauge parameter).  The first term on the 
right--hand side of Eq. (\ref{95}) corresponds to the ladder approximation.
The second (annihilation) term does not contribute to the BS
equation for NG bosons (this follows from the fact that, due
to the Ward identities for axial currents, the BS equation for NG
bosons can be reduced to the Schwinger-Dyson equation for the fermion
propagator where there is no contribution of the annihilation term \cite{6}).
For this reason we shall omit this term in the following.  Then the BS
equation takes the form:
\begin{eqnarray}
\chi^\beta_{AB}(x,y;P)&=&-4\pi\alpha \int d^4x_1 d^4y_1
S_{AA_1}(x,x_1) \delta_{a_1a_2} \gamma^\mu_{n_1n_2}
\chi^\beta_{A_2B_2}(x_1,y_1;P) \nonumber\\
&\times& \delta_{b_2b_1}\gamma^\nu_{m_2m_1}S_{B_1B}(y_1,y)
{\cal{D}}_{\mu\nu}(y_1-x_1)~, \label{97}
\end{eqnarray}
\no where, since the lowest in $\alpha$ (ladder) approximation is 
used, the full fermion propagator $G_{AB}(x,y)$ is replaced by the
propagator $S$ of a free fermion (with the mass $m=m_{\rm dyn}$) in
a magnetic field (see Eqs. (\ref{13}), (\ref{14})).

Using the new variables, the center of mass coordinate $R={x+y\over 2}$, 
and the relative coordinate $r=x-y$, Eq. (\ref{97}) can be rewritten as
\begin{eqnarray}
&&\tilde\chi_{nm}(R,r;P)=-4\pi\alpha \int d^4R_1d^4r_1 
\tilde S_{nn_1}\biggl(R-R_1+{r-r_1\over 2}\biggr)\nonumber\\
&&\times\gamma^\mu_{n_1n_2}\tilde\chi_{n_2m_2}(R_1,r_1;P)
\gamma^\nu_{m_2m_1} 
\tilde S_{m_1m}\biggl({r-r_1\over 2}-R+R_1\biggr)
{\cal{D}}_{\mu\nu}(-r_1)\nonumber\\
&&\times\exp\bigl[-ie(r+r_1)^\mu A_\mu^{\rm ext}(R-R_1)\bigr]  
\times \exp\bigl[iP(R-R_1)\bigr]~.  \label{98}
\end{eqnarray}
\no Here $\tilde S$ is defined in Eqs. (\ref{13}) and (\ref{14}), and the
function $\tilde\chi_{nm}(R,r;P)$ is defined from the equation
\begin{eqnarray}
&&\chi^\beta_{AB}(x,y;P) \equiv
\langle0|T\psi_A(x)\bar\psi_B(y)|P,\beta\rangle\nonumber\\
&&=\lambda^\beta_{ab}e^{-iPR} \exp\bigl[ ier^\mu A_\mu^
{\rm ext}(R)\bigr]\tilde\chi_{nm}(R,r;P) \label{99}
\end{eqnarray}
\no where $\lambda^\beta$ are $N_f^2-1$ flavor matrices
(${\rm tr}(\lambda^\beta \lambda^\gamma)=2\delta_{\beta\gamma};~
\beta,\gamma \equiv 1,\ldots,N_f^2-1$).  The important fact is that,
like in the case of the BS equation for the $\pi$ in the NJL model
considered in the previous section, the effect of translation
symmetry breaking by the magnetic field is factorized in the
Schwinger phase factor in Eq. (\ref{99}), and Eq. (\ref{98}) admits a translation
invariant solution, $\tilde\chi_{nm}(R,r;P)=\tilde\chi(r;P)$.  Then,
transforming this equation into momentum space, we get
\begin{eqnarray}
\tilde\chi_{nm}(p;P)&=&-4\pi\alpha \int
{d^2q_{\perp} d^2R_{\perp} d^2k_{\perp} d^2k_{\parallel}\over
(2\pi)^6} \nonumber\\
&\times& \exp\bigl[i({\bf P}_{\perp}-{\bf q}_{\perp}){\bf R}_{\perp}\bigr]
\tilde S_{nn_1}\biggl(p_{\parallel}+{P_{\parallel}\over 2},{\bf p}_{\perp}
+e{\bf A}^{\rm ext}({\bf R}_{\perp})
+{{\bf q}_{\perp}\over 2}\biggr) \nonumber\\
&\times& \gamma^\mu_{n_1n_2}\tilde\chi_{n_2m_2}(k,P)
\gamma^\nu_{m_2m_1}\tilde S_{m_1m}
\biggl(p_{\parallel}-{P_{\parallel}\over 2},
{\bf p}_{\perp}+e{\bf A}^{\rm ext}({\bf R}_{\perp})-
{{\bf q}_{\perp}\over 2}\biggr) \nonumber\\
&\times& {\cal{D}}_{\mu\nu}\bigl(k_{\parallel}-p_{\parallel},
{\bf k}_{\perp}-{\bf p}_{\perp}
-2e{\bf A}^{\rm ext}({\bf R}_{\perp})\bigr) \label{100}
\end{eqnarray}
\no (we recall that $p_{\parallel}\equiv (p^0,p^4)$,
${\bf p}_{\perp} \equiv (p^1,p^2)$). Henceforth we will consider
the equation with the total momentum $P_\mu\to 0$.

The crucial point for further analysis will be the assumption that
$m_{\rm dyn} << \sqrt{|eB|}$ and that the region mostly responsible
for generating the mass is the infrared region with 
$k \lsim m_{\rm dyn} \ll \sqrt{|eB|}$.  As we shall see, this
assumption is self-consistent (see Eq. (\ref{6})).  The assumption allows
us to replace the propagator $\tilde S_{nm}$ in Eq. (\ref{100}) by the
pole contribution of the LLL (see Eq. (\ref{79})), and Eq. (\ref{100}) transforms
into the following one:
\begin{eqnarray}
\rho(p_{\parallel},{\bf p}_{\perp}) &=&
{2\alpha\ell^2\over (2\pi)^4} e^{-\ell^2{\bf p}_{\perp}^2} \int
d^2A_{\perp} d^2k_{\perp} d^2k_{\parallel}
e^{-\ell^2{\bf A}_{\perp}^2}(1-i\gamma^1\gamma^2)\gamma^\mu \nonumber\\
&&\hspace{-25mm}\times {\hat k_{\parallel}+m_{\rm dyn}\over
k^2_{\parallel}-m^2_{\rm dyn}}
\rho(k_{\parallel},{\bf k}_{\perp})
{\hat k_{\parallel}+m_{\rm dyn}\over k^2_{\parallel}-m^2_{\rm dyn}}
\gamma^\nu(1-i\gamma^1\gamma^2) {\cal{D}}_{\mu\nu}(k_{\parallel}-p_{\parallel},
{\bf k}_{\perp}-{\bf A}_{\perp})~,     \label{101}
\end{eqnarray}
\no where $\rho(p_{\parallel},{\bf p}_{\perp})=
(\hat p_{\parallel}-m_{\rm dyn})\tilde\chi(p)
(\hat p_{\parallel}-m_{\rm dyn})$.  Eq. (\ref{101}) implies that
$\rho(p_{\parallel},{\bf p}_{\perp})=\exp(-\ell^2 {\bf p}_{\perp}^2)
\varphi(p_{\parallel})$, where $\varphi(p_{\parallel})$ satisfies
the equation
\begin{eqnarray}
\varphi(p_{\parallel}) &=&
{\pi\alpha\over (2\pi)^4} \int d^2k_{\parallel}
(1-i\gamma^1\gamma^2)\gamma^\mu
{\hat k_{\parallel}+m_{\rm dyn}\over k^2_{\parallel}-m^2_{\rm dyn}}
\varphi(k_{\parallel}) \nonumber\\
&\times& {\hat k_{\parallel}+m_{\rm dyn}\over k^2_{\parallel}-
m^2_{\rm dyn}} \gamma^\nu(1-i\gamma^1  \gamma^2) D^{\parallel}_{\mu\nu}
(k_{\parallel}-p_{\parallel})~. \label{102}
\end{eqnarray}
\no Here
\begin{eqnarray}
D^{\parallel}_{\mu\nu}(k_{\parallel}-p_{\parallel})=
\int d^2k_{\perp} \exp\biggl(-{\ell^2 {\bf k}^2_{\perp}\over 2}\biggr)
{\cal{D}}_{\mu\nu}(k_{\parallel}-p_{\parallel},{\bf k}_{\perp})~.
\label{103}
\end{eqnarray}
\no Thus, as in the NJL model in the previous section,
 the BS equation has been reduced to a two--dimensional
integral equation.  

Henceforth we will use Euclidean space with $k_4=-ik^0$. 
Then, because of the
symmetry $SO(2)\times SO(2)\times {\cal{P}}$ in a magnetic field, we
arrive at the matrix structure (\ref{84}) for $\varphi(p_{\parallel})$:
\begin{eqnarray}
\varphi(p_{\parallel})=\gamma_5(A+i\gamma_1\gamma_2 B+
\hat p_{\parallel}C+i\gamma_1\gamma_2 \hat p_{\parallel} D)
\label{104a}
\end{eqnarray}
\no where $A,B,C$ and $D$ are functions of $p^2_{\parallel}$.

We begin the analysis of Eq. (\ref{102}) by choosing the Feynman gauge (the
general covariant gauge will be considered below).  Then
\begin{eqnarray}
D^{\parallel}_{\mu\nu}(k_{\parallel}-p_{\parallel})=
i\delta_{\mu\nu} \pi \int^\infty_0
{dx \exp(-\ell^2 x/2)\over (k_{\parallel}-p_{\parallel})^2+x}~,
\label{104}
\end{eqnarray}
\no and, substituting the expression (\ref{104a}) for $\varphi(p_{\parallel})$
into Eq. (\ref{102}), we find that $B=-A$, $C=D=0$, i.e.
\begin{eqnarray}
\varphi(p_{\parallel})=A\gamma_5(1-i\gamma_1\gamma_2)~, \label{105}
\end{eqnarray}
\no and the function $A$ satisfies the equation:
\begin{eqnarray}
A(p)={\alpha\over 2\pi^2}\int {d^2kA(k)\over k^2+m^2_{\rm dyn}}
\int^\infty_0
{dx \exp(-\ell^2x/2)\over ({\bf k}-{\bf p})^2+x} \label{106}
\end{eqnarray}
(henceforth we will omit the symbol $\parallel$; for the explanation
of physical reasons of the appearance of the projection operator
$(1-i\gamma_1\gamma_2)/2$ in $\varphi(p)$, see footnote \ref{spin}).

Now introducing the function
\begin{eqnarray}
\Psi({\bf r}) = \int {d^2k\over (2\pi)^2}
{A(k)\over k^2+m^2_{\rm dyn}} e^{i{\bf kr}}~, \label{107}
\end{eqnarray}
\no we get the two--dimensional Schr\"odinger equation from Eq.
(\ref{106})
\begin{eqnarray}
\bigl(-\Delta+m^2_{\rm dyn}+V({\bf r})\bigr) \Psi({\bf r})=0~,
\label{108}
\end{eqnarray}
\no where the potential $V({\bf r})$ is
\begin{eqnarray}
V({\bf r})&=&-{\alpha\over 2\pi^2} \int d^2p e^{i{\bf pr}}
\int^\infty_0 {dx \exp(-x/2)\over \ell^2p^2+x}
=-{\alpha\over \pi\ell^2} \int^\infty_0 
dx e^{-x/2} K_0\biggl({r\over \ell} \sqrt{x}\biggr) \nonumber\\
&=& {\alpha\over \pi\ell^2} \exp\biggl({r^2\over 2\ell^2}\biggr)
Ei\biggl(-{r^2\over 2\ell^2}\biggr)         \label{109}
\end{eqnarray}
($K_0$ is the Bessel function).  The essential difference of the
potential (\ref{109}) with respect to the $\delta$-like potential
(\ref{89}) in
the NJL model is that it is long range.  Indeed, using the
asymptotic relations for $Ei(x)$ \cite{5}, we get:
\begin{eqnarray} 
V({\bf r}) &\simeq& -{2\alpha\over\pi}{1\over r^2}~, \quad
r\to \infty~, \nonumber\\
V({\bf r}) &\simeq& -{\alpha\over \pi\ell^2}
\biggl(\gamma+\ln {2\ell^2\over r^2}\biggr)~, \quad
r\to 0 ~,               \label{110}
\end{eqnarray}
\no where $\gamma\simeq 0.577$ is the Euler constant.  Therefore,
the theorem of Ref.\cite{7} (asserting that, for short-range potentials,
$-E(\alpha)\equiv m^2_{\rm dyn}(\alpha) \sim \exp(-1/a\alpha)$, with
$a>0$, as $\alpha\to 0$) cannot be applied to this case.  In
order to find $m^2_{\rm dyn}(\alpha)$, we shall use the integral
equation (\ref{106}). This equation is analyzed in Appendix C by using both
analytical and numerical methods. The result is:
\begin{eqnarray}
m_{\rm dyn}=C\sqrt{|eB|}\exp\biggl[-{\pi\over2}\biggl(\frac{\pi}{2\alpha}
\biggr)^{1/2}\biggr],\label{111}
\end{eqnarray}
where the constant $C=O(1)$. 
Note that this result agrees with the analysis of
Ref. \cite{19} where the analytic properties of $E(\alpha)$ were studied
for the Schr\"odinger equation with potentials having the asymptotics
$V({\bf r}) \to {\rm const}/r^2$ as $r\to\infty$.

Since
\begin{eqnarray}
\lim_{\alpha\to 0}{\exp\bigl[-1/a\sqrt{\alpha}\bigr]\over
{\exp\bigl[-1/a^\prime \alpha}\bigr]}=\infty~, \label{112}
\end{eqnarray}
\no at $a,a^\prime>0$, we see that the long-range character of the
potential leads to the essential enhancement of the dynamical mass.

Let us now turn to considering the general covariant gauge (\ref{96}).  As
is known, the ladder approximation is not gauge invariant.  However,
let us show that because the present effect is due to the infrared
dynamics in QED, where the coupling constant is small, the leading term
in $\ln(m_{\rm dyn}^2\ell^2)$, 
$\ln(m_{\rm dyn}^2\ell^2) \simeq -\pi\sqrt{\pi/2\alpha}$, 
is the same in all covariant gauges.

Acting in the same way as before, we find that the wave function
$\varphi(p)$ now takes the form
\begin{eqnarray}
\varphi(p)=\gamma_5(1-i\gamma^1\gamma^2) \bigl(A(p^2)+\hat pC(p^2)\bigr)   
\label{113}
\end{eqnarray}
\no where the functions $A(p^2)$ and $C(p^2)$ satisfy the equations:
\begin{eqnarray}
A(p^2)&=&{\alpha\over 2\pi^2} \int {d^2kA(k^2)\over
k^2+m^2_{\rm dyn}} \int^\infty_0 
{dx(1-\lambda x\ell^2/4)\exp(-x\ell^2/2)\over
({\bf k}-{\bf p})^2+x}~,  \label{114}\\
C(p^2) &=& {\alpha\lambda\over 4\pi^2} \int {d^2kC(k^2)\over
k^2+m^2_{\rm dyn}}\biggl[2k^2-({\bf kp})-{k^2({\bf kp})\over p^2}\biggr] 
\nonumber\\
&\times& \int^\infty_0
{dx\exp(-x\ell^2/2)\over \bigl[({\bf k}-{\bf p})^2+x\bigr]^2}~. 
\label{115}
\end{eqnarray}
\no One can see that the dominant contribution on the right-hand side of
Eq. (\ref{114}) (proportional to $[\ln m^2_{\rm dyn}\ell^2]^2$ and formed
at small $k^2$) is independent of the gauge parameter $\lambda$.  
Thus the leading contribution in $\ln(m_{\rm dyn}^2\ell^2)$, 
$\ln(m_{\rm dyn}^2\ell^2) \simeq -\pi\sqrt{\pi/2\alpha}$, 
is indeed gauge--invariant (for more details see Appendix C).

This concludes the derivation of Eqs. (\ref{6}), (\ref{108}) and
(\ref{109}) describing
spontaneous chiral symmetry breaking by a magnetic field in ladder QED.

Note the following important point. Because of the exponent 
$\exp(-\ell^2x/2)$ in integral equations (\ref{106}), (\ref{114})
and (\ref{115}), the present effect is connected with the infrared
dynamics in QED: the natural cutoff in this problem is $|eB|$. The
fact that the non--pertubative infrared dynamics in this problem
decouples from the ultraviolet dynamics is reflected also in the
asymptotic behavior of the functions $A(p^2)$ and $C(p^2)$: as
follows from Eqs. (\ref{106}), (\ref{114})
and (\ref{115}), these functions rapidly ($A(p^2) \sim 1/p^2$, 
$C(p^2) \sim \lambda /p^2$) decrease as $p^2\to\infty$. Therefore,
since at $p^2\gg |eB|$ the magnetic field essentially does not affect the
behavior of the running coupling in QED \cite{20}, the coupling
constant $\alpha$ in equations (\ref{106}), (\ref{114})
and (\ref{115}) has to be interpreted as the value of the running
coupling related to the scale $\mu^2\sim |eB|$.

Up to now we have considered the ladder approximation in QED in 
a magnetic field. As was already mentioned above, because of the
$(1+1)$--dimensional form of the fermion propagator in the infrared
region, there may be also relevant higher order contributions. 
In particular, the dimensional reduction may essentially affect the
vacuum polarization, thus changing the behavior of the running
coupling in the infrared region. We will show that this is indeed 
the case. Actually, in this problem, because a magnetic field breaks
Lorentz and $SO(3)$ rotational symmetries, it is more appropriate to
speak not about a single running coupling but about a
running coupling tensor, {\em i.e.} about the full photon propagator $\cal
D_{\mu\nu}$ in the magnetic field. 

Let us  consider the photon propagator in a strong magnetic field
in the infrared region ($|eB|\gg m^2_{\rm dyn}$, 
$|k^2_{\parallel}|$, $|k^2_{\perp}|$), with the
polarization operator calculated in one--loop approximation
\cite{20,21}.
One can rewrite it in the following form:
\begin{eqnarray}
{\cal D}_{\mu\nu}&=&-i\Bigg(\frac{1}{k^2}g_{\mu\nu}^{\perp}+
\frac{k_\mu^{\parallel}k_\nu^{\parallel}}{k^2k^2_{\parallel}}+
\nonumber\\
&+&\frac{1}{k^2+k^2_{\parallel}\Pi(k^2_{\parallel})}
(g^{\parallel}_{\mu\nu}-\frac{k^{\parallel}_{\mu}
k^{\parallel}_{\nu}}{k^2_{\parallel}})
-\frac{\lambda}{k^2}\frac{k_\mu k_\nu}{k^2}\Bigg),\label{116}\\
\Pi(k^2_{\parallel})&=&-\frac{\alpha}{2\pi}\frac{|eB|}{m^2_{\rm dyn}}
\Bigg[\frac{4m^2_{\rm dyn}}{k^2_{\parallel}}
-\frac{8m^4_{\rm dyn}}{k^2_{\parallel}\sqrt{(k^2_{\parallel})^2
-4m^2_{\rm dyn}k^2_{\parallel}}}\times\nonumber\\
&&\times\ln\frac{
\sqrt{
(k^2_{\parallel})^2-4m^2_{\rm dyn}
k^2_{\parallel}}-k^2_{\parallel}}
{\sqrt{(k^2_{\parallel})^2-4m^2_{\rm dyn}
k^2_{\parallel}}+k^2_{\parallel}}\Bigg],\label{117}
\end{eqnarray}
where the symbols $\perp$ and $\parallel$ in $g_{\mu\nu}$ are 
related to the  $(1,2)$ and $(0,3)$ components, respectively.
The asymptotic behavior of $\Pi(k^2_{\parallel})$ is:
\begin{eqnarray}
\Pi(k^2_{\parallel})&\to&\frac{\alpha}{3\pi}\frac{|eB|}{m^2_{\rm dyn}}
\qquad \mbox{at}\qquad |k^2_{\parallel}|\ll m^2_{\rm dyn},\label{118}\\
\Pi(k^2_{\parallel})&\to&-\frac{2\alpha}{\pi}\frac{|eB|}{k^2_{\parallel}}
\qquad \mbox{at}\qquad |k^2_{\parallel}|\gg m^2_{\rm dyn}. \label{119}
\end{eqnarray} 
Note the following characteristic points:

a) The expressions (\ref{116}), (\ref{117}) correspond to the one--loop
contribution with the fermions from the LLL. Actually, expression
(\ref{117}) is the leading term in the $1/|eB|$ expansion of the
one--loop polarization operator.

b) The polarization effects are absent in the transverse components
of ${\cal D}_{\mu\nu}$. This is because the spin of fermions from the
LLL is polarized along the magnetic field (see footnote \ref{spin}). Indeed,
this implies that in the QED--vertex, with two fermions from the LLL, 
the photon spin equals zero along the magnetic field, {\em i.e.} only
the longitudinal, $(0,3)$, components, are relevant in this case.

c) There is a strong screening effect in the $(g^{\parallel}_{\mu\nu}-
k^{\parallel}_\mu k^{\parallel}_{\nu}/k^2_{\parallel})$--component
of the photon propagator. In particular there is a pole at
$k^2_{\parallel}=0$ in 
$\Pi(k^2_{\parallel})$ as $m^2_{\rm dyn}\to 0$: this is a
reminiscence of the Higgs effect in the $(1+1)$--dimensional massless
QED (Schwinger model) \cite{22}.

d) Is the dynamics in QED in a magnetic field similar to that in the 
Schwinger model, as $|eB|\to \infty$? The answer to this question 
is ``no". The crucial point is that, unlike the NJL model, 
there is a neutral fundamental 
field in the QED Lagrangian: the photon field. As one
can see from Eq.(\ref{116}), the photon propagator does not have a
$(1+1)$--dimensional form even as $|eB|\to\infty$: there is the
quantity $k^2=k^2_0-k^2_{\perp}-k^2_{3}$ in the denominator of the
photon propagator. This leads to the interaction which  is very
different from that in the Schwinger model. 
Notice that the interaction at long distances in this problem 
is much weaker than that in
the $(1+1)$--dimensional Schwinger model: while here, in Euclidean
space, the potential $V(r)$  is $V(r)\sim-1/r^2$ as $r\to\infty$ 
(see Eq.(\ref{110})), it is $V(r)\sim \ln{r}$ in the Schwinger model.
Thus the infrared dynamics in QED in a magnetic field is a complex
mixture of $(3+1)$ and  $(1+1)$--dimensional dynamics. 

Let us now consider the BS equation in the so called improved ladder
approximation: in this approximation, the free photon propagator is
replaced by the one--loop propagator (\ref{116}). As was already
indicated above, the transverse component of the photon propagator
decouple from the fermions at the LLL. The longitudinal components
lead to the following equation (in the Feynman gauge): 
\begin{eqnarray}
A(p^2_{\parallel})&=&\frac{\alpha}{4\pi^2}\int 
\frac{d^2k_{\parallel}A(k^2_{\parallel})}{k^2_{\parallel}+m^2_{\rm dyn}}
\int\limits^\infty_0dx e^{-\ell^2x/2}\bigg[
\frac{1}{x+(k_{\parallel}-p_{\parallel})^2}+\nonumber\\
&+&\frac{1}{x+(k_{\parallel}-p_{\parallel})^2
+(k_{\parallel}-p_{\parallel})^2\Pi_{E}(k_{\parallel}-p_{\parallel})}
\bigg], 
\label{120a}
\end{eqnarray}
where $\Pi_{E}$ is the polarization operator (\ref{117}) in
Euclidean space (compare with Eq.(\ref{106})). The first term in the
square brackets on the right--hand side of Eq.(\ref{120a}) comes from
the (unscreened) $k^{\parallel}_\mu
k^{\parallel}_{\nu}/k^2_{\parallel}$--component; the second term
comes from the (screened) $(g^{\parallel}_{\mu\nu}-
k^{\parallel}_\mu k^{\parallel}_{\nu}/k^2_{\parallel})$--component.
It is not difficult to show that at $p^2=0$ the latter does not
contribute to the dominant term (proportional to 
$[\ln{m_{\rm dyn}^2\ell^2}]^2$) on the right--hand side of this
equation. Indeed, one finds
\begin{eqnarray}
A(0)\sim\frac{\alpha}{4\pi^2}A(0)\int 
\frac{d^2k_{\parallel}}{k^2_{\parallel}+m^2_{\rm dyn}}
\int\limits^\infty_0dx e^{-\ell^2x/2}\bigg[
\frac{1}{x+k_{\parallel}^2}
+\frac{1}{x+k_{\parallel}^2+k_{\parallel}^2\Pi_{E}(k_{\parallel})}\bigg], 
\label{121a}
\end{eqnarray}
and, matching asymptotics (\ref{118}), (\ref{119}) at
$k_{\parallel}^2=6m^2_{\rm dyn}$ in Euclidean space, 
we get the following estimate for the second term:
\begin{eqnarray}
&~&\frac{\alpha}{4\pi^2}A(0)\int 
\frac{d^2k_{\parallel}}{k^2_{\parallel}+m^2_{\rm dyn}}
\int\limits^\infty_0dx e^{-\ell^2x/2}
\frac{1}{x+k_{\parallel}^2+k_{\parallel}^2\Pi_{E}(k_{\parallel})}\sim
\nonumber\\
&\sim&\frac{\alpha}{4\pi}A(0)\int\limits^\infty_0dx e^{-\ell^2x/2}\bigg[
\int\limits^{6m^2_{\rm dyn}}_0 \frac{dy}{(y+m^2_{\rm
dyn})[x+y(1+\alpha/3\pi m^2_{\rm dyn}\ell^2)]}
+\nonumber\\
&&+\int\limits_{6m^2_{\rm dyn}}^\infty \frac{dy}{(y+m^2_{\rm dyn})
(x+y+2\alpha/\pi\ell^2)}\bigg]\simeq\nonumber\\
&\simeq&\frac{\alpha}{4\pi}A(0)\ln\bigg(\frac{2}{7m^2_{\rm dyn}\ell^2}
\bigg)\ln\frac{\pi}{\alpha}.
\label{122a}
\end{eqnarray}
Therefore the dominant contribution in this equation is connected 
with the unscreened $k^{\parallel}_\mu 
k^{\parallel}_{\nu}/k^2k^2_{\parallel}$ 
component (since $k^{\parallel}_\mu$ is different from $k_{\mu}$, this
term is not a gauge artifact wich can be removed by a gauge
transformation). As a result, we get the same equation as in the
ladder approximation but with $\alpha$ replaced by $\alpha/2$. The
expression for $m_{\rm dyn}$ is now given by Eq.(\ref{111}) with
$\alpha\to\alpha/2$. It is gauge invariant. Also, since $\alpha$ is
the value of the running coupling at a definite scale, $\mu^2\sim
|eB|$, this expression is renormalization group invariant.

The present consideration shows that, despite the smallness of $\alpha$,
the expansion  in $\alpha$ is broken in the infrared region in this
model. There are two reasons for that: the $(1+1)$--dimensional
character of the fermion propagator in the infrared region and the
smallness of the dynamical mass $m_{\rm dyn}$ as compared to
$|eB|^{1/2}$ ($\ln(|eB|/m_{\rm dyn}^2)\sim 1/\sqrt{\alpha}$). 

It is well known that in massive QED, infrared singularities (on 
mass shell) in Green's functions can be completely factorized if
a certain set of diagrams is summed up \cite{8}. The situation in
massless QED is much more complicated: the set of relevant diagrams
becomes essentially larger and, up to now, the problem of the
complete description of the infrared singularities in that model 
remains unsolved. The smallness of $m_{\rm dyn}$ in QED 
in a magnetic field makes this
model closer to massless QED than to massive one. This implies that,
despite the smallness of the coupling constant $\alpha$, the infrared
dynamics in this model is quite complicated.

It is a challenge to define the class of all those diagrams in QED in
a magnetic field which give a relevant contribution in this 
problem. Since the QED coupling
constant is weak in the infrared region, this problem, though
hard, seems not to be hopeless.

\ve

\ce {\bf 10. CONCLUSION}
\b

In this paper we showed that a constant magnetic field is a strong
catalyst of spontaneous chiral symmetry breaking in 3+1 and 2+1
dimensions, leading to the generation of a fermion dynamical mass 
even at the weakest attractive interaction between fermions.  The 
essence of this effect is the dimensional reduction $D\to D-2$ 
in the dynamics of fermion pairing in a magnetic field.  In particular, 
the dynamics of NG modes, connected with spontaneous chiral symmetry 
breaking by a magnetic field is described by the two (one)--dimensional 
Schr\"odinger equation at $D=3+1~(D=2+1)$:
\begin{eqnarray}
\bigl(-\Delta+m^2_{\rm dyn}+V({\bf r})\bigr)
\Psi({\bf r})=0~, \label{120}
\end{eqnarray}
\no where the attractive potential is model dependent and it
defines the form of the dynamical mass as a function of the
coupling constant.  Since the general theorem of Ref. \cite{7} assures the
existence of at least one bound state for the two- and one--dimensional
Schr\"odinger equations with an attractive potential, the generation of
the dynamical mass $m_{\rm dyn}$ takes place even at the weakest
attractive interaction between fermions at $D=3+1$ and $D=2+1$.  This
general effect was illustrated in the NJL model and QED.

In this paper, we considered the dynamics in the presence of a
constant magnetic field only.  It would be interesting to extend
this analysis to the case of inhomogeneous electromagnetic fields.
In connection with this, note that
in 2+1 dimensions, the present effect is intimately connected with the
fact that the massless Dirac equation in a constant magnetic field admits
an infinite number of normalized solutions with $E=0$ (zero modes) \cite{2}.
More precisely, the density of the zero modes 
\begin{eqnarray*}
\tilde\nu_0 = \lim_{S\to\infty} S^{-1}N(E)\biggl|_{E=0}
\end{eqnarray*}
\no (where $S$ is a two--dimensional volume of the system) is finite.
As has been already pointed out (see the second paper in Ref. \cite{2} and
Ref. \cite{23}), spontaneous flavor (chiral) symmetry breaking in 2+1
dimensions should be catalysed by all stationary (i.e. independent 
of time) field configurations with $\tilde\nu_0$ being finite.  On the
other hand, as we saw in this paper, the density
\begin{eqnarray*}
\nu_0=\lim_{V\to\infty} V^{-1} {dN(E)\over dE}\Biggl|_{E=0}
\end{eqnarray*}
\no of the states with $E=0$ (from a continuous spectrum) plays the crucial
role in the catalysis of chiral symmetry breaking in 3+1 dimensions.
One may expect that the density $\nu_0$ should play an important role
also in the case of (stationary) inhomogeneous configurations in
3+1 dimensions.

As a first step in studying this problem, it would be important to 
extend the Schwinger results \cite{1} to inhomogeneous field configurations.
Interesting results in this direction have been recently obtained in
Ref. \cite{24}.

In conclusion, let us discuss possible applications of this effect.

Since $(2+1)$--dimensional relativistic field theories may serve as
effective theories for the description of long wavelength excitations
in planar condensed matter systems \cite{25}, this effect may be 
relevant for such systems.  It would be also interesting to take
into account this effect in studying the possibility of
the generation of a magnetic field in the vacuum, i.e. spontaneous 
breakdown of the Lorentz symmetry, in $(2+1)$--dimensional QED \cite{26}.

In 3+1 dimensions, one potential application of the effect can be
connected with the possibility of the existence of very strong
magnetic fields ($B\sim 10^{24}G$) during the electroweak phase
transition in the early universe \cite{27}.  As the results obtained in
this paper suggest, such fields might essentially change the
character of the electroweak phase transition.

Another application of this effect can be connected with the role of
chromomagnetic backgrounds as models for the QCD vacuum (the
Copenhagen vacuum \cite{28}).  

Yet another potentially interesting application
is the interpretation of the results of the GSI heavy-ion scattering
experiments in which narrow peaks are seen in the energy spectra of
emitted $e^+e^-$ pairs \cite{29}.  One proposed explanation \cite{30} is that
a strong electromagnetic field, created by the heavy ions, induces a
phase transition in QED to a phase with spontaneous chiral symmetry
breaking and the observed peaks are due to the decay of positronium-like
states in the phase.  The catalysis of chiral symmetry breaking by a
magnetic field in QED, studied in this paper, can serve as a toy
example of such a phenomenon.  In order to get a more realistic model,
it would be interesting to extend this analysis to non-constant background
fields \cite{31}.

We believe that the effect of the dimensional reduction by external
field configurations may be quite general and relevant for 
different non-perturbative phenomena.  It deserves further study.
\ve

\ce {\bf ACKNOWLEDGMENTS}
\b

V.P.G. is grateful to the members of the Institute for Theoretical
Physics of the University of Groningen, especially D. Atkinson, for
their hospitality. He wishes to acknowledge the Stichting FOM (Fundamenteel
Onderzoek der Materie), financially supported by the Nederlandse Organisatie
voor Wetenschappelijk Onderzoek, for its support.  V. A. M. thanks the 
members of the Department of Physics of the University of California, Los 
Angeles, particularly D. Cangemi and E. D'Hoker, for their hospitality during 
his stay at UCLA. We are grateful to G. McKeon and T. Sherry for
useful discussions. The work of I.A.Sh. was supported in part by the 
International Science Education Program (ISSEP) through Grant No. PSU052143.

\ve
\ce {\bf APPENDIX A}
\b

In this Appendix, we derive the kinetic term ${\cal{L}}_k$ in the 
effective action $\Gamma$ (\ref{34}) in the NJL model.

The structure of the kinetic term (\ref{63}) is:
\begin{eqnarray}
{\cal{L}}={F_1^{\mu\nu}\over 2}(\partial_\mu\rho_j \partial_\nu\rho_j)
+{F_2^{\mu\nu}\over \rho^2} (\rho_j\partial_\mu\rho_j)
(\rho_i\partial_\nu\rho_i)~, \label{121}
\end{eqnarray}
\no where ${\mbox{\boldmath$\rho$}}=(\sigma,\pi)$ and $F_1^{\mu\nu},
F_2^{\mu\nu}$ depend on the $U_{\rm L}(1)\times U_{\rm R}(1)$-invariant
$\rho^2=\sigma^2+\pi^2$.  The definition $\Gamma=\int d^4x{\cal{L}}$
and Eq. (\ref{121}) imply that the functions $F_1^{\mu\nu}$, $F_2^{\mu\nu}$
are determined from the equations:
\begin{eqnarray}
{\delta^2\Gamma_k\over \delta\sigma(x) \delta\sigma(0)}\Biggl|_
{\scriptstyle \sigma=const\atop\scriptstyle\pi=0}
&=&-(F_1^{\mu\nu}+2F_2^{\mu\nu})\Biggl|_
{\scriptstyle\sigma=const\atop\scriptstyle\pi=0}
\cdot \partial_\mu\partial_\nu\delta^4(x)~,    \label{122}\\
{\delta^2\Gamma_k\over\delta\pi(x)\delta\pi(0)}\Biggl|_
{\scriptstyle\sigma=const\atop\scriptstyle\pi=0}
&=&-F_1^{\mu\nu}\Biggl|_
{\scriptstyle\sigma=const\atop\scriptstyle\pi=0}
\cdot \partial_\mu\partial_\nu\delta^4(x)~.   \label{123}
\end{eqnarray}
\no Here $\Gamma_k$ is the part of the effective action containing
terms with two derivatives.  Eq. (\ref{34}) implies that $\Gamma_k=\tilde\Gamma_k$.
Therefore we find from Eq. (\ref{123}) that
\begin{eqnarray}
F_1^{\mu\nu}=-{1\over 2} \int d^4xx^\mu x^\nu
{\delta^2\tilde\Gamma_k\over \delta\pi(x)\delta\pi(0)} =
-{1\over 2} \int d^4xx^\mu x^\nu
{\delta^2\tilde\Gamma\over \delta\pi(x)\delta\pi(0)}    \label{124}
\end{eqnarray}
(henceforth we shall not write explicitly the condition
$\sigma$=const., $\pi=0$).  Taking into account the definition of the
fermion propagator,
\begin{eqnarray}
iS^{-1}=i\hat D-\sigma~,               \label{125}
\end{eqnarray}
\no we find from Eq. (\ref{36}):
\begin{eqnarray}
{\delta^2\tilde\Gamma\over 
\delta\pi(x)\delta\pi(0)}&=&-i~{\rm tr}\bigl(S(x,0)
i\gamma^5 S(0,x)i\gamma^5\bigr) \nonumber\\
&=& -i~{\rm tr}\bigl(\tilde S(x)i\gamma^5 \tilde S(-x)i\gamma^5\bigr) 
\nonumber\\
&=& -i \int {d^4kd^4q\over (2\pi)^8}
e^{iqx}{\rm tr}\bigl(\tilde S(k)i\gamma^5
\tilde S(k+q)i\gamma^5\bigr)      \label{126}
\end{eqnarray}
\no (the functions $\tilde S(x)$ and $\tilde S(k)$ are given in
Eqs. (\ref{13}) and (\ref{14}) with $m=\sigma$).  
Therefore
\begin{eqnarray}
F_1^{\mu\nu}=-{i\over 2} \int {d^4k\over (2\pi)^4} {\rm tr}
\biggl(\tilde S(k) i\gamma^5
{\partial^2\tilde S(k)\over \partial k_\mu \partial k_\nu}
i\gamma^5\biggr)~.                    \label{127}
\end{eqnarray}
\no In the same way, we find that
\begin{eqnarray}
F_2^{\mu\nu}=-{i\over 4} \int {d^4k\over (2\pi)^4} {\rm tr}
\biggl(\tilde S(k){\partial^2 \tilde S(k)\over
\partial k_\mu\partial k_\nu}-\tilde S(k) i\gamma^5
{\partial^2\tilde S(k)\over\partial k_\mu\partial k_\nu}
i\gamma^5\biggr)~.                        \label{128}
\end{eqnarray}
\no Taking into account the expression for $\tilde S(k)$ in Eq.
(\ref{14})
(with $m=\sigma$), we get:
\begin{eqnarray}
{\partial^2\tilde S(k)\over\partial k_0\partial k_0}&=&
2iN_c\ell^4 \int^\infty_0 dt\cdot te^{R(t)}
\bigl[\bigl(2it(\ell k^0\bigr)^2
[k^0\gamma^0-k^3\gamma^3+\sigma] \nonumber\\
&&+3k^0\gamma^0-k^3\gamma^3+\sigma\bigr)
(1+\eta\gamma^1\gamma^2T)-{\bf k}_{\perp}{\mbox{\boldmath $\gamma$}}_{\perp}
(1+2i(\ell k^0)^2t)\nonumber\\
&&\times (1+T^2)\bigr]~,        \label{129}\\
{\delta^2\tilde S(k)\over \partial k_3\partial k_3}&=&
-2iN_c\ell^4 \int^\infty_0 dt\cdot te^{R(t)}
\bigl[(-2it(\ell k^3)^2[k^0\gamma^0-k^3\gamma^3+\sigma] \nonumber\\
&&+k^0\gamma^0-3k^3\gamma^3+\sigma)(1+\eta\gamma^1\gamma^2T)
-{\bf k}_{\perp}{\mbox{\boldmath $\gamma$}}_{\perp}
(1-2i(\ell k^3)^2t)\nonumber\\
&&\times(1+T^2)\bigr]~,                \label{130}\\
{\partial^2\tilde S(k)\over \partial k_j \partial k_j}&=&
-2iN_c\ell^4 \int^\infty_0 dt\cdot Te^{R(t)}
\bigl[(-2iT(\ell k^j)^2[k^0\gamma^0-k^3\gamma^3+\sigma\bigr] \nonumber\\
&&+k^0\gamma^0-k^3\gamma^3+\sigma)(1+\eta\gamma^1\gamma^2T)
-{\bf k}_{\perp}{\mbox{\boldmath $\gamma$}}_{\perp}(1-2i(\ell k^j)^2T)
(1+T^2)\nonumber\\
&&-2k^j\gamma^j(1+T^2)\bigr]~,   \label{131}
\end{eqnarray}
\no where $T=\tan t$, $\eta={\rm sign}(eB)$, $R(t)=i\ell^2t
\bigl((k^0)^2-(k^3)^2-\sigma^2\bigr)-i\ell^2{\bf k}^2_{\perp}T$,
and $j=1,2$ (there is no summation over $j$).

Eqs. (\ref{14}), (\ref{127}), and (\ref{128}) imply that non-diagonal terms
$F_1^{\mu\nu}$ and $F_2^{\mu\nu}$ are equal to zero.  Below, we
will consider in detail the calculation of the function $F_1^{00}$;
other functions $F_1^{\mu\mu}$, $F_2^{\nu\nu}$ can similarly be found.

Substituting expressions (\ref{14}) and (\ref{129}) into Eq.
(\ref{127}) for $F_1^{00}$,
we get
\begin{eqnarray}
F_1^{00} &=& {N_c\ell^2\over 4\pi^4} \int d^4k \int^\infty_0
\int^\infty_0 d\tau dt e^{(R(t)+R(\tau))} tA\bigl[2i(\ell k^0)^2 t
\bigl((k^0)^2-(k^3)^2-\sigma^2\bigr) \nonumber\\
&+&3(k^0)^2-(k^3)^2-\sigma^2\bigr)
-B{\bf k}_{\perp}^2(1+2i(\ell k^0)^2t)\bigr]~, \label{132}
\end{eqnarray}
\no where $A=1-T{\cal{T}}$, $B=(1+T^2)(1+{\cal{T}}^2)$,
${\cal{T}}=\tan \tau$.  After integrating over $k$, we find:
\begin{eqnarray}
F_1^{00} &=& {N_c\over 4\pi^2} \int^\infty_0 \int^\infty_0
{d\tau dt\over (\tau+t)^2}\tau t 
e^{-i\ell^2\sigma^2(t+\tau)} \nonumber\\
&\times& \biggl[{A\over T+{\cal{T}}}
\biggl({2\over \tau+t}+i\ell^2\sigma^2\biggr)
+{B\over (T+{\cal{T}})^2} \biggr]~.      \label{133}
\end{eqnarray}
\no Changing $t$ and $\tau$ to new variables,
\begin{eqnarray}
t={s\over 2}(1+u)~, \quad  \tau={s\over 2}(1-u)~,
\label{134}
\end{eqnarray}
\no and then introducing the imaginary (dimensionless) proper time,
$s\to -is$, we arrive at the expression:
\begin{eqnarray}
F_1^{00} &=& {N_c\over 24\pi^2} \int^\infty_0
{ds\over s} e^{-(\ell^2\sigma^2s)}
\biggl[(2s\coth s-2)+\biggl({s^2\over \sinh^2s}-1\biggr)
+\ell^2\sigma^2s^2\coth s\biggr] \nonumber\\
&+& {1\over 8\pi^2} \int^\infty_{1/\Lambda^2}{ds\over s}
e^{-(\ell^2\sigma^2s)} \nonumber\\
&=& {N_c\over 8\pi^2} \biggl[\ln {\Lambda^2\ell^2\over 2}-\psi
\biggl({\sigma^2\ell^2\over 2}+1\biggr)+
{1\over \sigma^2\ell^2}-\gamma+{1\over 3}\biggr]~. 
\end{eqnarray}
\no Here we used relation (\ref{46}) and the relations \cite{5}:
\begin{eqnarray}
\int^\infty_0 e^{-\beta x}
\biggl({1\over x}-\coth x\biggr) dx=\psi
\biggl(1+{\beta\over 2}\biggr)-\ln{\beta\over 2}-{1\over \beta}~; \quad
{\rm Re}~\beta>0~, \label{136}
\end{eqnarray}
\begin{eqnarray}
\int^\infty_0{x^{\mu-1}e^{-\beta\tau}\over \sinh^2\tau}
d\tau &=&
2^{1-\mu}\Gamma(\mu)
\biggl[2\zeta\biggl(\mu-1,~{\beta\over 2}\biggr) \nonumber\\
&-&\beta\zeta\biggl(\mu,~{\beta\over 2}\biggr)\biggr]~; \quad
\mu > 2 
\end{eqnarray}
\no All the other functions $F_1^{\mu\mu},~ F_2^{\mu\mu}$ can similarly 
be found.
\ve

\ce {\bf APPENDIX B}
\b

In this Appendix we analyze the next--to--leading order in $1/N_c$
expansion in the $(3+1)$--dimensional NJL model (the analysis of the
$(2+1)$--dimensional NJL model is similar).  Our main goal is to show
that the propagator of the neutral NG boson $\pi$ has a 
$(3+1)$--dimensional form in this approximation and that (unlike the
$(1+1)$--dimensional Gross--Neveu model \cite{13}), the $1/N_c$ expansion is
reliable in this model.

For an excellent review of the $1/N_c$ expansion see Ref. \cite{32}.  For
our purposes, it is sufficient to know that the perturbative
expansion is given by Feynman diagrams with the vertices and the
propagators of fermions and composite particles $\sigma$ and $\pi$
calculated in leading order in $1/N_c$.  In the leading order, the
fermion propagator is given in Eqs. (\ref{13}) and (\ref{14}) 
(with $m$ replaced by $m_{\rm dyn}$).  As follows from Eq. 
(\ref{32}), the bare Yukawa coupling of fermions with $\sigma$ 
and $\pi$ is $g^{(0)}_{\rm Y}=1$ in this approximation.
The inverse propagators of $\sigma$ and $\pi$ are \cite{16}:
\begin{eqnarray}
D^{-1}_{\mbox{\boldmath$\rho$}}(x)=N_c
\biggl({\Lambda^2\over 4\pi^2g} \delta^4(x)+i~{\rm tr}
\bigl[S(x,0)T_{\mbox{\boldmath$\rho$}}S(0,x)T_{\mbox{\boldmath$\rho$}}
\bigr]\biggr)~,\label{138}
\end{eqnarray}
\no where ${\mbox{\boldmath$\rho$}}=(\sigma,\pi)$ and $T_\sigma =1$, $T_\pi =i\gamma^5$.  
Here $S(x,0)$ is the fermion propagator (\ref{13}) with the mass
$m_{\rm dyn}=\bar\sigma$ defined from the gap equation (\ref{48}).
Actually, for our purposes, we need to know the form of the 
propagator of $\pi$ at small momenta only.  We find from Eqs.
(\ref{63}) and (\ref{64}):
\begin{eqnarray}
D_\pi(k)=-{8\pi^2\over N_c} f_1(\Lambda \ell,\bar\sigma \ell)
\bigl[k_0^2-k_3^2-{\bf k}_{\perp}^2
f_2(\Lambda \ell,\bar\sigma \ell)\bigr]^{-1},
\label{139}
\end{eqnarray}
\no where
\begin{eqnarray}
f_1 &=& \biggl[\ln{\Lambda^2\ell^2\over 2}-\psi
\biggl({\bar\sigma^2\ell^2\over 2}+1\biggr)+
{1\over \bar\sigma^2\ell^2}-\gamma+{1\over 3}\biggr]^{-1}~, \\
f_2 &=& \biggl(\ln{\Lambda^2\over \bar\sigma^2}-\gamma+{1\over 3}
\biggr) f_1~. 
\end{eqnarray}
\no The crucial point for us is that, because of the dynamical mass
$m_{\rm dyn}$, the fermion propagator (despite its $(1+1)$--dimensional
form) is soft in the infrared region and that the propagator of $\pi$
has a $(3+1)$--dimensional form in the infrared region (as follows from
Eqs. (\ref{63}) and (\ref{64}), the propagator of $\sigma$ has of course also a
$(3+1)$--dimensional form).

Let us begin the analysis by considering the next--to--leading order
corrections in the effective potential.  The diagram which contributes 
to the effective potential in this order is shown in Fig. 1a.  Because 
of the structure of the propagators pointed out above, there are no
infrared divergences in this contribution to the potential.  (Note that
this is in contrast to the Gross--Neveu model: the contribution of this
diagram is logarithmically divergent in the infrared region in that
model, i.e. the $1/N_c$ expansion is unreliable in that case).  
Therefore the diagram in Fig.1a leads to a finite, $O(1)$, correction
to the potential $V$ (we recall that the leading contribution in $V$ is
of order $N_c$).  As a result, at sufficiently large values of $N_c$
the gap equation in this model admits a non--trivial solution
$\bar\sigma \not= 0$ in next--to--leading order in $1/N_c$, i.e. 
there is spontaneous chiral symmetry breaking in this approximation.

Let us now consider the next--to--leading order corrections to the
propagator of the NG mode $\pi$.  First of all, note that in a
constant magnetic field, the propagator of a neutral local field
$\varphi(x)$, $D_\varphi(x,y)$, is translation invariant, i.e. it
depends on $(x-y)$.  This immediately follows from the fact that the
operators of space translations in Eqs. (\ref{28}) and (\ref{29}) take the
canonical form for neutral fields (the operator of time translations
is $i{\partial\over \partial t}$ both for neutral and charged fields
in a constant magnetic field).  The diagram contributing to the
propagators of the NG mode in this order is shown in Fig. 1b.
Because of the dynamical mass $m_{\rm dyn}$ in the fermion propagator,
this contribution is analytic at $k_\mu=0$.  Since at large $N_c$ the
gap equation has a non-trivial solution in this approximation, there is
no contribution of $O(k^0)\sim {\rm const.}$ in the inverse
propagator of $\pi$.  Therefore the first term in the momentum expansion
in its inverse propagator has the form $C_1(k_0^2-k_3^2)
-C_2 {\bf k}_{\perp}^2$, where $C_1$ and $C_2$ are functions of
$\bar\sigma\ell$ and $\Lambda\ell$ and the propagator takes the following
form in this approximation:
\begin{eqnarray}
D_\pi(k) =_{_{_{_{\hspace{-4mm}{{k\to}{ 0}}}}}}
&-&{8\pi^2\over N_c} f_1(\Lambda\ell,\bar\sigma\ell)
\biggl[\biggl(1-{1\over N}\tilde C_1
(\Lambda\ell,\bar\sigma \ell)\biggr)(k_0^2-k_3^2) \nonumber\\
&-&\biggl(f_2(\Lambda\ell,\bar\sigma\ell)-{1\over N} \tilde C_2
(\Lambda\ell,\bar\sigma\ell)\biggr) {\bf k}^2_{\perp}\biggr]^{-1} 
\end{eqnarray}
\no (compare with Eq. (\ref{139})).

For the same reasons, there are also no infrared divergences in 
the fermion propagator (see Fig.1c) nor in the Yukawa
vertices (see Fig.1d) in this order.  Therefore, at sufficiently
large values of $N_c$, the results retain essentially the same as in
leading order in $1/N_c$.

We believe that there should be no essential obstacles to extend
this analysis for all orders in $1/N_c$.
\ve

\ce{\bf APPENDIX C}
\b

In this Appendix we analyze integral equations (\ref{106}) and
(\ref{114}) for the
function $A(p)$. We will show that these equations lead to the expression
(\ref{111}) for $m_{\rm dyn}$.

The integral equations were analyzed by using both analytical and numerical
methods \footnote{We thank Anthony Hams and Manuel Reenders for their help
in numerical solving these equations.}.
The numerical plot of the dynamical mass as a function of $\alpha$ is shown
in Fig.2. Below we consider the analytical analysis of these equations.

We begin by considering equation (\ref{106}) in the Feynman gauge.
After the angular integration, this equation becomes
\begin{eqnarray}
A(p^2)=\frac{\alpha}{2\pi}\int\limits_0^\infty\frac{dk^2A(k^2)}
{k^2+m^2_{\rm dyn}}K(p^2,k^2)\label{143}
\end{eqnarray}                
with the kernel
\begin{eqnarray}
K(p^2,k^2)=\int\limits_0^\infty\frac{dz\exp(-z\ell^2/2)}{\sqrt{
(k^2+p^2+z)^2-4k^2p^2}}.\label{144}
\end{eqnarray}
To study Eq.(\ref{143}) analytically it is convenient to break the momentum
integration into two regions and expand the kernel appropriately for 
each region (compare with Ref.\cite{11}):

\begin{eqnarray}
A(p^2)&=&\frac{\alpha}{2\pi}\biggl[\int\limits_0^{p^2}\frac{dk^2A(k^2)}
{k^2+m^2_{\rm dyn}}\int\limits_0^\infty\frac{dz\exp(-z\ell^2/2)}
{p^2+z}\nonumber\\
&+&\int\limits_{p^2}^\infty\frac{dk^2A(k^2)}{k^2+m^2_{\rm dyn}}\int
\limits_0^\infty\frac{dz\exp(-z\ell^2/2)}{k^2+z}\biggr].\label{145}
\end{eqnarray}
Introducing dimensionless variables $x=p^2\ell^2/2,\,y=k^2\ell^2/2,\,
a=m^2_{\rm dyn}\ell^2/2$, we rewrite Eq.(\ref{145}) in the form
\begin{eqnarray}
A(x)=\frac{\alpha}{2\pi}\biggl[g(x)\int\limits_0^{x}\frac{dyA(y)}
{y+a^2}+\int\limits_x^\infty\frac{dyA(y)g(y)}{y+a^2}\biggr],\label{146}
\end{eqnarray}   
where
\begin{eqnarray}
g(x)=\int\limits_0^\infty\frac{dze^{-z}}{z+x}=-e^xEi(-x),\label{147}
\end{eqnarray}
and $Ei(x)$ is the integral exponential function. The solutions of integral 
equation (\ref{146}) satisfy the second order differential equation
\begin{eqnarray}
A^{\prime\prime}-\frac{g^{\prime\prime}}{g^\prime}A^\prime-\frac{\alpha}
{2\pi}g^\prime\frac{A}{x+a^2}=0,\label{148}
\end{eqnarray}
where $(\prime)$ denotes derivative with respect to $x$. The boundary
conditions to this equation follow from the integral equation
(\ref{146}):
\begin{eqnarray}
&&\frac{A^\prime}{g^\prime}\biggl|_{x=0},   \label{149}\\
&&\bigl(A-\frac{gA^\prime}{g^\prime}\bigr)\biggl|_{x=\infty}=0.
\label{150}
\end{eqnarray}
For the derivatives of the function $g(x)$ we have relations
\begin{eqnarray}
g^\prime=-{1\over x}+g(z),\quad g^{\prime\prime}={1\over x^2}-{1\over x}
+g(x),\label{151}
\end{eqnarray}
and $g(x)$ has asymptotics
\begin{eqnarray}
g(x)&\sim& \ln{e^{-\gamma}\over x},\quad x\to 0,\nonumber\\
g(x)&\sim&{1\over x}-{1\over x^2}+{2\over x^3},\quad
x\to\infty.\label{152}
\end{eqnarray}

Using Eqs.(\ref{151}),(\ref{152}) we find that Eq.(\ref{148}) has two independent solutions 
which behave as $A(x)\sim const$, $A(x)\sim\ln{1/x}$ and $A(x)\sim const$, 
$A(x)\sim {1\over x}$, near $x=0$ and $x=\infty$, respectively.
The infrared boundary condition (BC) (\ref{149}) leaves only the solution with 
regular behaviour, $A(x)\sim const$, while the ultraviolet BC gives an 
equation to determine $a=a(\alpha)$.
To find analytically $a(\alpha)$ we will solve approximate equations in
regions $x<<1$ and $x>>1$ and then match the solutions at the point $x=1$.
This provides insight into the critical behaviour of the solution at
$\alpha\to 0$. A numerical study of
the full Eq.(\ref{143}) reveals the same approach to criticality (see
the plot in Fig.2).

In the region $x<<1$ Eq.(\ref{148}) is reduced to a hypergeometric type equation:
\begin{eqnarray}
A^{\prime\prime}+{1\over x}A^\prime+\frac{\alpha}{2\pi}\frac{A}{x(x+
a^2)}=0.\label{153}
\end{eqnarray}
The regular at $x=0$ solution has the form
\begin{eqnarray}
A_1(x)=C_1F(i\nu,-i\nu,1;-{x\over a^2}),\quad \nu=\sqrt{\alpha
\over 2\pi},\label{154}
\end{eqnarray}
where $F$ is a hypergeometric function \cite{5}.
In the region $x>>1$ Eq.(\ref{148}) takes the form
\begin{eqnarray}
A^{\prime\prime}+{2\over x}A^\prime+\frac{\alpha}{2\pi}\frac
{A}{x^2(x+a^2)}=0.\label{155}
\end{eqnarray}
The solution satisfying ultraviolet BC (\ref{150}) is
\begin{eqnarray}
A_2(x)=C_2{1\over x} F({1+i\mu\over2},{1-i\mu\over2};2;-{a^2\over x}),
\quad\mu=\sqrt{\frac{2\alpha}{\pi a^2}-1}.\label{156}
\end{eqnarray}
Equating now logarithmic derivatives of $A_1$ and $A_2$ at $x=1$ we
arrive at the equation determing the quantity $a(\alpha)$:
\begin{eqnarray}
\frac{d}{dx}\biggl\{\ln\frac{xF(i\nu,-i\nu;1;-{x\over a^2})}{F({1+i\mu
\over2},{1-i\mu\over2};2;-{a^2\over
x})}\biggr\}\biggl|_{x=1}=0.\label{157}
\end{eqnarray}
Note that up to now we have not made any assumptions on the value of
the parameter $a$. Let us seek now for solutions of Eq.(\ref{157}) with $a<<1$
(which corresponds to the assumption of the LLL dominance).
Then the hypergeometric function in denominator of Eq.(\ref{157}) can be replaced
by 1 and we are left with equation
\begin{eqnarray}
-{1\over a^2}\nu^2F(1+i\nu,1-i\nu;2;-{1\over a^2})+F(i\nu,-i\nu;1;-{1
\over a^2})=0,\label{158}
\end{eqnarray}
where we used the formula for differentiating the hypergeometric
function \cite{5}
\begin{equation}
\frac{d}{dz}F(a,b,c;z)=\frac{ab}{c}F(a+1,b+1,c+1;z).\label{159}
\end{equation}
Now, because of $a<<1$, we can use the formula of asymptotic
behavior of hypergeometric function at large values  of its argument
$z$ \cite{5}:
\begin{equation}
F(a,b,c;z)\sim\frac{\Gamma(c)\Gamma(b-a)}{\Gamma(b)\Gamma(c-a)}(-z)^{-a}
+\frac{\Gamma(c)\Gamma(a-b)}{\Gamma(a)\Gamma(c-b)}(-z)^{-b}.\label{160}
\end{equation}
Then Eq.(\ref{158}) is reduced to the following one:
\begin{eqnarray}
&&\cos\biggl[\nu\ln{1\over a^2}+\arg\Sigma(\nu)\biggr]=0,\nonumber\\
&&\Sigma(\nu)={1+i\nu\over2}\frac{\Gamma(1+2i\nu)}{\Gamma^2(1+i\nu)}
\label{161}
\end{eqnarray}
and we get
\begin{eqnarray}
m^2_{\rm dyn}=2|eB|\exp\biggl[-\frac{\pi(2n+1)/2-\arg\Sigma(\nu)}
{\nu}\biggr],\label{162}
\end{eqnarray}
where $n$ is zero or positive integer.
The $\arg\Sigma(\nu)$ can be rewritten as
\begin{eqnarray}
\arg\Sigma(\nu)=\arctan\nu+\arg\Gamma(1+2i\nu)-2\arg\Gamma(1+i\nu)
\label{163}
\end{eqnarray}
and in the limit $\nu\to 0$ Eq.(\ref{157}) takes the form
\begin{eqnarray}
m^2_{\rm dyn}=2|eB|\mbox{e}\exp\biggl[-\frac{\pi}{2\nu}(2n+1)\biggr]=2|eB|
\mbox{e}\exp\biggl[-\pi\sqrt{\pi\over2\alpha}(2n+1)\biggr]\label{164}
\end{eqnarray}
(the second factor $\mbox{e}$ here is $\mbox{e}\simeq 2.718$ and not the 
electric charge!).
The stable vacuum corresponds to the largest value of $m^2_{\rm dyn}$
with $n=0$.

Let us now turn to equation (\ref{114}) in an arbitrary covariant gauge. The
function $g(x)$ is now replaced by
\begin{eqnarray}
\tilde g(x)=\int\limits_0^\infty\frac{dze^{-z}(1-\lambda z/2)}{z+x}=g(x)+
{1\over2}\lambda xg^\prime(x).\label{165}
\end{eqnarray}
As it is easy to verify, this does not change equation (\ref{153}) in
the region $x<<1$. At $x>>1$ we have
\begin{eqnarray}
\tilde g^\prime(x)\sim
-\frac{2-\lambda}{2x^2}+2\frac{1-\lambda}{x^3},\nonumber\\
\frac{\tilde g^{\prime\prime}(x)}{\tilde g^\prime(x)}=-{2\over x}\frac
{2-\lambda-6{1-\lambda\over x}}{2-\lambda-4{1-\lambda\over
x}}.\label{166}
\end{eqnarray}
Therefore in any gauge, except $\lambda=2$, the differentional equation 
at $x>>1$ takes the form
\begin{eqnarray}
A^{\prime\prime}+{2\over x}A^\prime+\frac{\alpha(2-\lambda)}{4\pi}\frac
{A}{x^2(x+a^2)}=0 \label{167}
\end{eqnarray}
with asymptotic solution $A(x)\sim{1\over x}$. In the gauge $\lambda=2$,
instead of Eq.(\ref{167}), we have
\begin{eqnarray}
A^{\prime\prime}+{3\over x}A^\prime
+\frac{\alpha}{\pi}\frac{A}{x^3(x+a^2)}=0,\label{168}
\end{eqnarray}
which gives more rapidly decreasing behavior $A(x)\sim{1\over x^2}$ when
$x\to\infty$. Repeating the previous analysis, we are led to 
expression (\ref{111}) for $m_{\rm dyn}$.

\ve

\ve

\center{FIGURE CAPTIONS}
\bigskip
\begin{enumerate}
\item{}{Fig. 1.}  Diagrams in next-to-leading order in $1/N_c$.  A solid
line denotes the fermion propagator and a dashed line denotes the
propagators of $\sigma$ and $\pi$ in leading order in $1/N_c$.
\item{}{Fig. 2.} A numerical fit of the dynamical mass as a 
function of the coupling constant $\alpha$: $m_{\rm
dyn}/\sqrt{2eB}=\exp[-\frac{\pi}{2}\sqrt{\frac{\pi}{2\alpha}}a+b]$. 
The fitting parameters are: $a=1.000059$, $b= -0.283847$.
\end{enumerate}

\vfill\eject
\begin{figure}
\epsfxsize=12cm
\epsffile[100 400 400 766]{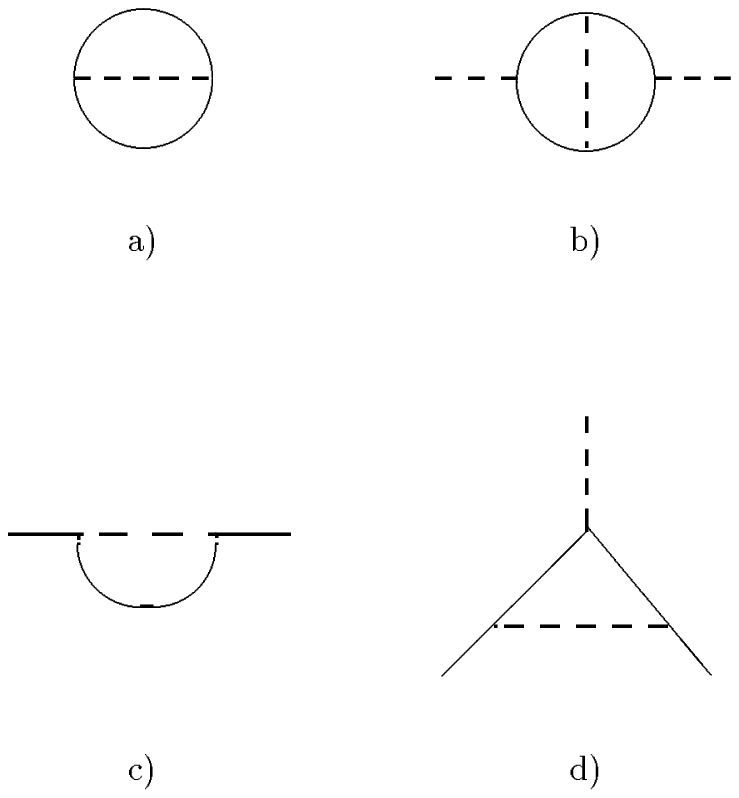}
\end{figure}
\vspace{1cm}

Figure 1: Diagrams in next-to-leading order in $1/N_c$.  A solid
line denotes the fermion propagator and a dashed line denotes the
propagators of $\sigma$ and $\pi$ in leading order in $1/N_c$.

\vfill\eject
\begin{figure}
\label{fig2}
\epsfxsize=12cm
\epsffile[100 400 400 766]{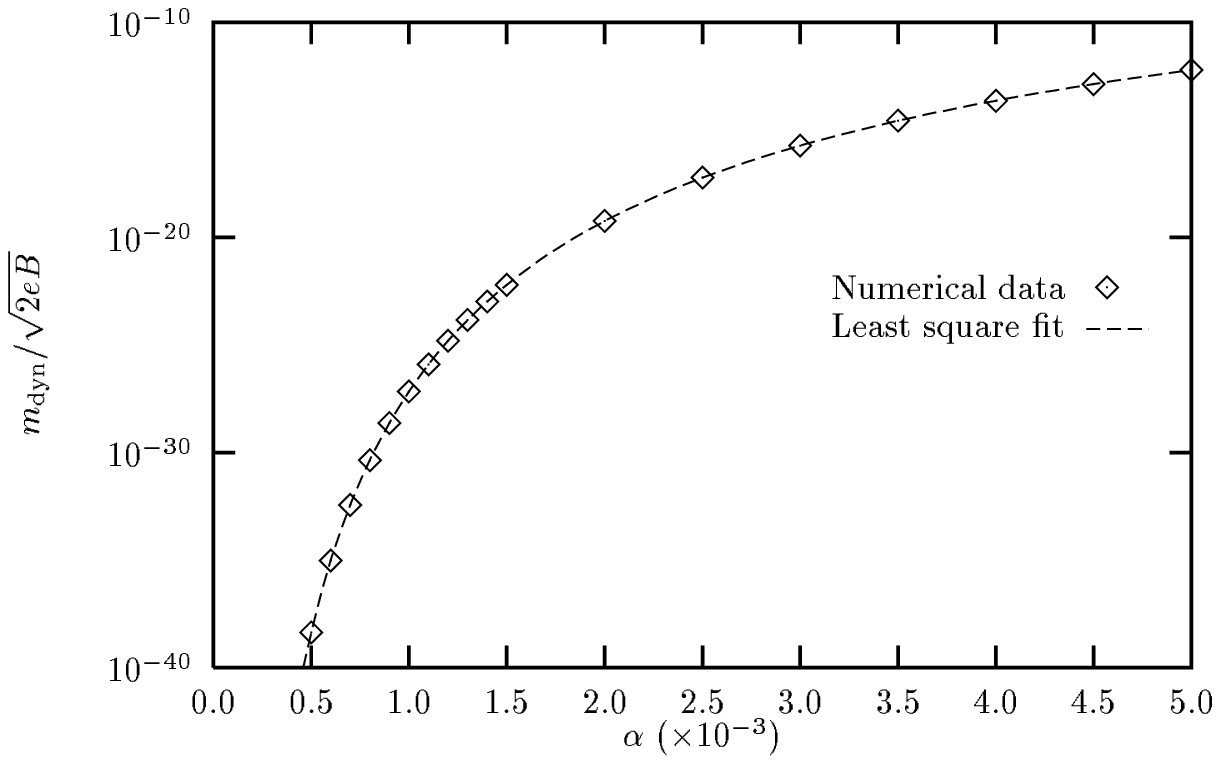}
\end{figure}
\vspace{1cm}

Figure 2: A numerical fit of the dynamical mass as a 
function of the coupling constant $\alpha$: $m_{\rm
dyn}/\sqrt{2eB}=\exp[-\frac{\pi}{2}\sqrt{\frac{\pi}{2\alpha}}a+b]$. 
The fitting parameters are: $a=1.000059$, $b= -0.283847$.

\end{document}